\documentclass[conf]{new-aiaa}
\usepackage[utf8]{inputenc}

\usepackage{graphicx}
\usepackage{amsmath}
\usepackage[version=4]{mhchem}
\usepackage{siunitx}
\usepackage{longtable,tabularx}
\usepackage[labelformat=empty]{subfig}
\usepackage{booktabs}
\usepackage{breqn}
\usepackage{bm}
\usepackage{multirow}
\usepackage{pdflscape}
\usepackage{siunitx}
\usepackage{color}

\newcommand{\ppartial}[2]{\dfrac{\partial #1}{\partial #2}}

\makeatletter
\renewcommand*\env@matrix[1][\arraystretch]{%
  \edef\arraystretch{#1}%
  \hskip -\arraycolsep
  \let\@ifnextchar\new@ifnextchar
  \array{*\c@MaxMatrixCols c}}
\makeatother

\title{Probabilistic Flight Envelope Estimation with Application to Unstable Over-Actuated Aircraft}

\author{Mingzhou Yin\footnote{MSc Student, Department of Control and Simulation, Faculty of Aerospace Engineering, Kluyverweg 1; mingzhouyin@gmail.com.}, Q. P. Chu\footnote{Associate Professor, Department of Control and Simulation, Faculty of Aerospace Engineering, Kluyverweg 1; q.p.chu@tudelft.nl.}, and Y. Zhang\footnote{Ph.D. Student, Department of Control and Simulation, Faculty of Aerospace Engineering, Kluyverweg 1; y.zhang-9@tudelft.nl.}}
\affil{Delft University of Technology, 2629 HS Delft, The Netherlands}

\author{Michael A. Niestroy\footnote{Aeronautical Eng. Manager, Lockheed Martin ADP, P.O. Box 748, Ft Worth, TX 76101, AIAA Associate Fellow.}}
\affil{Lockheed Martin Aero, Fort Worth, TX, 76101}

\author{C. C. de Visser\footnote{Assistant Professor, Department of Control and Simulation, Faculty of Aerospace Engineering, Kluyverweg 1; c.c.devisser@tudelft.nl.}}
\affil{Delft University of Technology, 2629 HS Delft, The Netherlands}

\begin{document}

\maketitle

\begin{abstract}
This paper proposes a novel and practical framework for safe flight envelope estimation and protection, in order to prevent loss-of-control-related accidents. Conventional analytical envelope estimation methods fail to function efficiently for systems with high dimensionality and complex dynamics, which is often the case for high-fidelity aircraft models. In this way, this paper develops a probabilistic envelope estimation method based on Monte Carlo simulation. This method generates a probabilistic estimation of the flight envelope by simulating flight trajectories with extreme control effectiveness. It is shown that this method can significantly reduce the computational load compared with previous optimization-based methods and guarantee feasible and conservative envelope estimation of no less than seven dimensions. This method was applied to the Innovative Control Effectors aircraft, an over-actuated tailless fighter aircraft with complex aerodynamic coupling between control effectors. The estimated probabilistic flight envelope is used for online envelope protection by a database approach. Both conventional state-constraint-based and novel predictive probabilistic flight envelope protection systems were implemented on a multi-loop nonlinear dynamic inversion controller. Real-time simulation results demonstrate that the proposed framework can protect the aircraft within the estimated envelope and save the aircraft from maneuvers that otherwise would result in loss of control.
\end{abstract}

\section*{Nomenclature}

{\renewcommand\arraystretch{1.0}
\noindent\begin{longtable*}{@{}l @{\quad=\quad} l@{}}
$A_x,A_y,A_z$		& linear accelerometer measurements, ft/s$^2$ \\
$b$					& wing span, ft \\
$\bar{c}$			& mean aerodynamic chord, ft \\
$d$					& number of dimensions \\
$E$ 				& dynamic flight envelope \\
$\tilde{E}$			& probabilistic dynamic flight envelope \\
$F_x,F_y,F_z$		& aerodynamic forces in the body frame, lbf \\
$\hat{f}(\bm{x})$	& kernel density estimator \\
$g$ 				& gravitational acceleration, ft/s$^2$ \\
$h$					& altitude, ft \\
$h_j$				& bandwidth of the $j$th variable \\
$I,R$ 				& invariant and reachable sets \\
$J$					& inertia matrix, slug$\cdot$ft$^2$ \\
$\tilde{J}$			& augmented inertia matrix \\
$\bm{J}_d$			& full control effectiveness matrix \\
$\bm{J}_d^F,\bm{J}_d^M$		& control effectiveness matrix of aerodynamic forces and moments \\
$\bm{J}_{env}$		& gradient of the probabilistic envelope metric \\
$K$ 				& trim set \\
$k(\cdot)$ 			& kernel function \\
$k_0$				& threshold setting of the probabilistic envelope \\
$l(\bm{x})$			& level set function of the trim set \\
$m$					& mass of the aircraft, slug \\
$M$					& Mach number \\
$\bm{M}_{env}$		& flight envelope metric \\
$M_x,M_y,M_z$		& aerodynamic moments in the body frame, ft$\cdot$lbf \\
$N$ 				& number of samples \\
$n,s$ 				& number of states and inputs \\
$p,q,r$				& roll, pitch, and yaw rate, rad/s \\
$\bar{q}$			& dynamic pressure, psi \\
$\tilde{R}$			& fuzzy reachable set \\
$S$					& total wing area, ft$^2$ \\
$S_\Phi$ 			& set of safe flight trajectories \\
$T$ 				& thrust, lbf \\
$T_f$ 				& time horizon, s \\
$U(t)$ 				& set of admissible input functions \\
$u,v,w$				& velocity components in the body frame, ft/s \\
$V(\cdot)$			& level set function of the invariant set \\
$V_g$				& ground speed, ft/s \\
$\bm{W}$			& control input sampling weights \\
$\bm{X}$			& random vector of states \\
$\bm{x}$ 			& state vector \\
$\bm{x}_e$ 			& effective states in envelope estimation \\
$\alpha,\beta$ 		& angle of attack and side-slip angle, rad \\
$\bm{\delta}$ 		& control input vector \\
$\eta$ 				& speedup factor \\
$\bm{\nu}$			& virtual control \\
$\bm{\nu}_h$		& hedged virtual control \\
$\sigma$			& standard deviation \\
$\bm{\Phi}$ 		& state trajectory \\
$\phi,\theta,\psi$ 	& roll, pitch, and yaw angle, rad \\
$\bm{\chi}$ 		& probabilistic envelope compensation term \\
\end{longtable*}}

\section{Introduction}
\label{sec:1}

\lettrine{S}{afety} is the most crucial issue in all sections of aviation, including flight control system (FCS) design. To reduce future accidents, it was observed by various sources (e.g.~\cite{RN19}) that loss of control (LOC) in flight is the most common primary cause of fatal accidents. LOC occurs when the aircraft has left the part of the state space where aircraft are safe to operate, which is commonly known as the safe flight envelope \cite{RN21}. With knowledge of the flight envelope available, LOC can be avoided by adequate envelope protection as shown by case study \cite{RN78}.

Conventionally, flight envelopes only deal with slow variables like altitude and airspeed in steady or quasi-steady conditions to achieve upset prevention. However, this type of envelope fails to take the dynamics into account and is usually determined empirically from flight tests \cite{RN30}. Therefore, a new type of envelope named the dynamic envelope \cite{RN28} or the immediate envelope \cite{RN36}, is defined as all the possible states where an airplane can both reach from and be controlled back to a set of initial flight conditions (usually trimmed) within a given recovery time. This definition of the safe flight envelope is used for the remainder of the paper.

Due to its critical position in safety, flight envelope estimation has been investigated extensively. Among different methods applied, the most rigorous and elaborately studied method is that based on reachability analysis, which formulates reachability problems within an optimal control framework by studying all possible controlled trajectories of dynamic systems \cite{RN39}. Various methods were developed to solve the reachability problem. For linear systems, convex optimization \cite{RN5} and geometric methods \cite{RN71} were applied based on the convexity of the flight envelope. For nonlinear systems, the distance-fields-over-grids method \cite{RN4} optimizes the state trajectories toward points in a predefined grid. The flight envelope can also be represented in the form of a level set function \cite{RN16, RN9, RN40} that solves the Hamilton-Jacobi-Bellman partial differential equation (HJB-PDE) resulting from an optimal control problem. These solutions are computed using the `level-set method', which is a widely-used numerical toolbox for solving nonlinear optimal control problems \cite{Mitchell_2000}.

However, the common problem with these methods to numerically solve the optimization problem and/or partial differential equations is the high computational load for complex nonlinear systems with high dimensions \cite{RN9}. The usual solution is to simply restrict investigation to problems with low dimensions by introducing virtual inputs with time-scale separation \cite{RN36,RN37} or perform domain decomposition \cite{RN85}. The maximum dimension implemented directly was four \cite{RN23}. 6-D estimation is possible when combined with system decomposition \cite{RN85}.

When the safe flight envelope has been estimated, flight envelope protection (FEP) systems strive to prevent LOC by constraining aircraft within the estimated envelope, preferably in both autopilot and manual control modes \cite{RN54}. FEP is often implemented to be a human-machine interface as a soft extension to the FCS by open-loop cueing to pilots \cite{RN56}. This includes stick shakers or pushers \cite{RN4} and special display design \cite{RN70}. Such design focuses on increasing pilots' situation awareness.

In contrast to the above concept, with the emergence of advanced flight control systems, it is more desirable to have FEP embedded in the controller itself, such that commands can be automatically justified when the aircraft is impaired or close to LOC. This concept targets at reduced workload, which is observed by both objective and subjective measures \cite{Lombaerts_2017}. Such a concept for FEP is used for the remainder of the paper. The inclusion of FEP has been studied in many types of controllers, including hybrid control \cite{RN54} and model-based predictive control \cite{RN55}. In fault-tolerant flight control, FEP can be included in the reconfiguring controller \cite{RN36}. Preliminary implementation of this idea has been observed on modern commercial aircraft such as Boeing 777 and Airbus A380 to avoid stalls or limit load factors. However, to the best of our knowledge, none of the work was conducted to apply systematic FEP to a multi-loop nonlinear controller.

In this paper, a probabilistic envelope estimation and protection framework is proposed that aims to fill the above gaps observed in previous research, namely high-dimensional envelope estimation for complex systems and FEP with multi-loop nonlinear control. The framework replaces the high-dimensional optimal reachability problem, which is computationally infeasible, with sub-optimal but feasible Monte Carlo (MC) estimation. To avoid severe underestimation from sub-optimality, an effective control input sampling strategy is proposed, dubbed the extreme control effectiveness method. This method examines the derived equation in the optimal control framework and only samples within a small population of candidate optimal controls. In this way, the envelope estimation problem is generally tractable even for high-dimensional and complex models with appropriate sample sizes.

In addition to the tractability problem, almost all previous research models the flight envelope as a crisp set. However, the actual ability to adopt an effective maneuver to save the aircraft depends on a range of practical issues such as external disturbances, pilots' reaction time, actuator dynamics, modeling errors, and controller performance. Hence, the safety of a given state is probabilistic. This probability can be characterized by the range of possible safe trajectories available. Therefore, this work extends the definition of the flight envelope to fuzzy sets with kernel density estimation, to include relative degree of safety.

The estimated envelope is protected in a multi-loop nonlinear dynamic inversion (NDI) controller. Despite enhanced computational efficiency, the estimation still cannot be conducted online. So a database approach is used, which stores the offline-estimated envelope for LOC avoidance. In fault-tolerant control, multiple pre-computed envelopes under different failure cases can be interpolated as in Ref.~\cite{RN86, RN87}.

With the crisp flight envelope definition, FEP is almost always implemented by limiting certain signals in the FCS. This can be achieved by direct state-command limiting, or indirect limiting of control surface deflections or virtual controls with a mapping from the envelope boundaries to avoid dependencies on specific controller structures \cite{RN55}. Such approaches are also applicable to the probabilistic envelope by binarizing it with a threshold setting. However, it is not the best practice since it makes the FCS completely 'blind' before the limits are activated. Therefore, with additional information provided by the probabilistic envelope, a novel predictive protection law is proposed to react earlier and introduce gentler protective actions. Instead of hard command limiting, this method modifies commands throughout a flight. The outer-loop commands are further modified by generalized multi-loop pseudo control hedging (PCH) to avoid future violation of the envelope.

\begin{figure}[hbt!]
\centering
\includegraphics[width=3.25in]{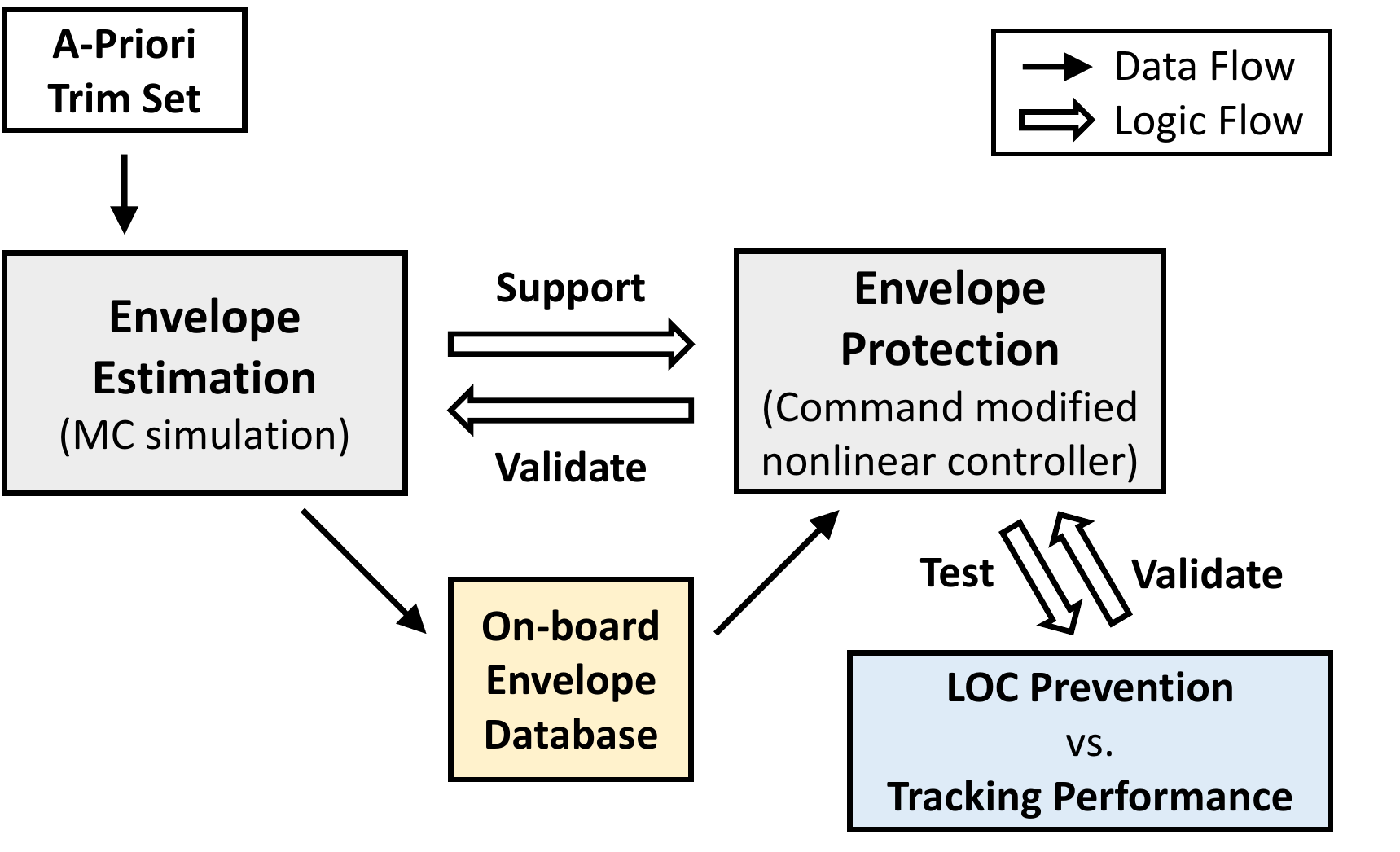}
\caption{General schematic of the proposed envelope estimation and protection framework.}
\label{fig:schematic}
\end{figure}

The general schematic of this framework is summarized in Fig.~\ref{fig:schematic}.

This framework is applied to a high-performance fighter aircraft, Lockheed Martin’s Innovative Control Effectors (ICE) aircraft \cite{RN22} to demonstrate its capability. The ICE aircraft is a tailless aircraft with a suite of 13 control effectors, including innovative ones such as spoiler-slot deflectors and multi-axis thrust vectoring. Together with improved lift-to-drag ratio and reduced weight, the ICE aircraft is highly maneuverable at high angle of attack (AoA). It is noted that the above challenges are more prominent on such aircraft, as 1) tailless aircraft are inherently unstable with less directional control authority, 2) with highly coupled aerodynamics associated with unconventional control effectors, the flight envelope tends to be nonlinear as well, and 3) extreme maneuvers are expected to be made which are closer to LOC. It was observed in previous research \cite{RN25} that the ICE aircraft is prone to LOC when controlled without FEP systems due to its extreme control authority. High-fidelity simulation shows by case study that both proposed FEP strategies can successfully avoid LOC with a seven-dimensional envelope database. To our best effort, this high-dimensional non-linear envelope estimation is not tractable by other existing methods.

The paper is organized as follows. Sec.~\ref{sec:2} introduces the reachability analysis framework for envelope estimation. The methodology to use MC simulation in probabilistic envelope estimation is presented in Sec.~\ref{sec:3}. Sec.~\ref{sec:4} further discusses the design of the envelope database and associated envelope protection laws. The designed FEP system is integrated into an NDI controller in Sec.~\ref{sec:5}. The implementation of the framework is demonstrated in Sec.~\ref{sec:7}, with a description of the applied ICE aircraft model. The envelope estimation results and controller performance comparisons are shown in Sec.~\ref{sec:8}. Sec.~\ref{sec:9} concludes the paper.

\section{The Reachability Analysis Formulation}
\label{sec:2}

\subsection{Classical Safe Flight Envelope}

It has been shown that the flight envelope estimation problem can be formulated as a reachability problem with the optimal control framework by studying possible trajectories of the dynamic system \cite{RN39}. Consider an autonomous nonlinear system

\begin{equation}
\dot{\bm{x}} = \bm{f}(\bm{x}, \bm{\delta}(t)),
\label{eq:forward}
\end{equation}

\noindent with $\bm{x}\in \mathbb{R}^n$, $\bm{\delta}(t)\in U(t)$, which defines the set of admissible input functions. Given a kernel set $K\in \mathbb{R}^n$ and a time horizon $T_f > 0$, the reachable set $R(T_f, K)$ and the invariant set $I(T_f, K)$ of the system are defined as follows.

\begin{equation}
R(T_f, K) = \left\{\bm{x}(0)\in \mathbb{R}^n 
\left\vert
\exists \bm{\delta}(t)\in U(t), \exists \tau \in [0,T_f], \bm{x}(\tau)\in K
\right.
\right\},
\label{eq:reach}
\end{equation}
\begin{equation}
I(T_f, K) = \left\{\bm{x}(0)\in \mathbb{R}^n
\left\vert
\forall \bm{\delta}(t)\in U(t), \forall \tau \in [0,T_f], \bm{x}(\tau)\in K
\right.
\right\}.
\label{eq:invariance}
\end{equation}

\begin{figure}[hbt!]
\centering
\includegraphics[width=0.5\linewidth]{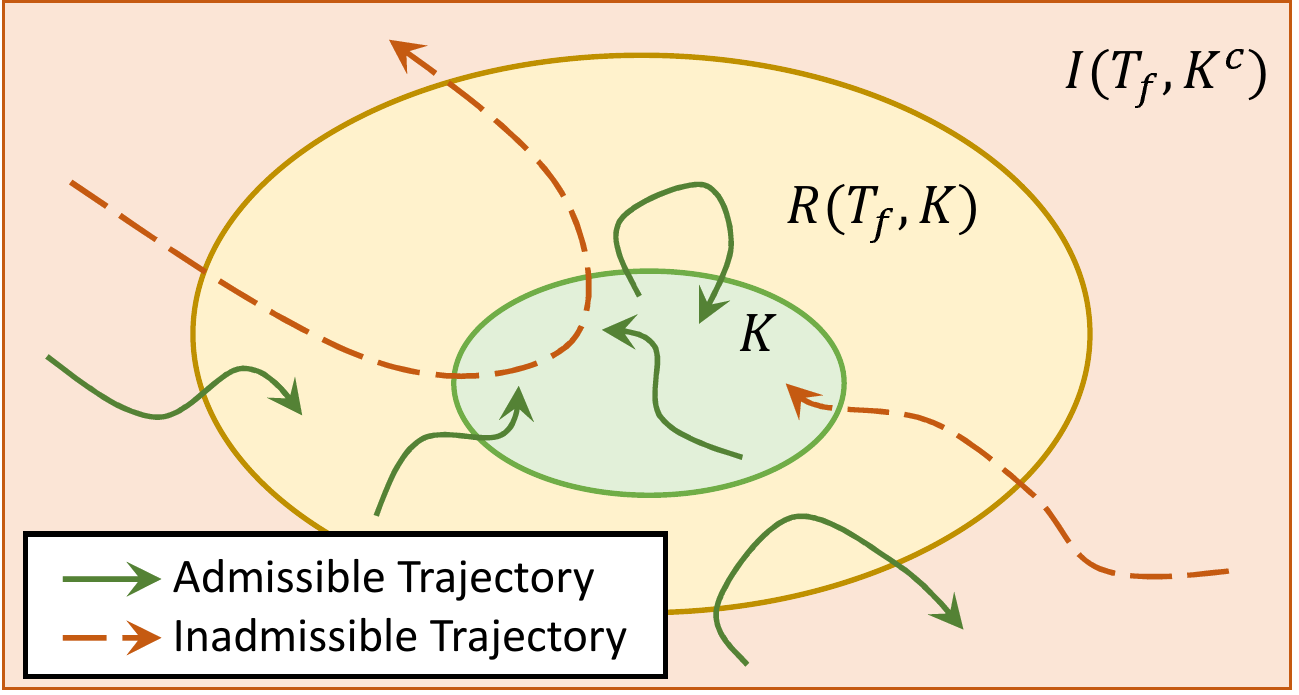}
\caption{Illustration of the duality between reachable sets and invariant sets.}
\label{fig:dual}
\end{figure}

\noindent Note that this definition uses fixed initial time at 0 and a variable horizon, which slightly differs from Ref.~\cite{RN39} where the terminal time is fixed. The reachable and the invariant sets are connected by the following principle of duality.

\begin{equation}
R(T_f, K) = (I(T_f, K^c))^c,
\label{eq:dual}
\end{equation}

\noindent where $(\cdot)^c$ denotes set complement. The advantage of expressing the reachable set $R(T_f,K)$ in terms of the invariant set $I(T_f, K^c)$ is that the invariant set can be computed directly. As shown in Fig.~\ref{fig:dual}, $I(T_f, K^c)$ consists of all admissible state trajectories that do not enter the trim set $K$ within a given time horizon. Clearly, the reachable set then consists of all states that do not belong to $I(T_f, K^c)$.

A similar definition can be applied to the same system, but running reversely in time,

\begin{equation}
\dot{\bm{x}} = -\bm{f}(\bm{x}, \bm{\delta}).
\label{eq:backward}
\end{equation}

\noindent In this way, the reachable set for the original system is known as the backward reachable set $R_b(T_f, K)$, whereas that for the reverse system is known as the forward reachable set $R_f(T_f, K)$.

In the context of flight envelope estimation, $K$ can be selected as the apriori trim set of the aircraft, where the aircraft can stay indefinitely. $T_f$ characterizes the available recovery time for pilots. Then, $R_f(T_f, K)$ defines all the states the aircraft can reach within $T_f$ from steady states, whereas $R_b(T_f, K)$ defines all the states from which the aircraft can be controlled back to steady states within $T_f$. In this way, the flight envelope to be estimated is defined as the intersection of the above two reachable sets 

\begin{equation}
E(T_f,K)=R_f(T_f,K)\cap R_b(T_f,K),
\label{eq:inter}
\end{equation}

\noindent according to the definition of the dynamic envelope.

\subsection{Reachability Analysis as an Optimal Control Problem}

Define the trim set as

\begin{equation}
K=\left\{\bm{x}\in \mathbb{R}^n \left\vert l(\bm{x}) > 0\right.\right\},
\label{eq:trimset}
\end{equation}

\noindent where $l(\bm{x}): \mathbb{R}^n \rightarrow \mathbb{R}$ represents a continuous cost function which is problem specific, see Ref.~\cite{RN39,RN16} for possible implementations. In Ref.~\cite{RN39}, the invariant set is expressed in terms of a level set function $V(\bm{x},T_f)$ sliced from a value function $V(\bm{x},t)$ as

\begin{equation}
I(T_f, K^c) = \left\{\bm{x}\in \mathbb{R}^n 
\left\vert V(\bm{x},T_f)\leq 0
\right.
\right\},
V(\bm{x},t)=\inf_{\bm{\delta}(\cdot)\in U(\cdot)} \min_{\tau\in[0,t]} l(\bm{\Phi}(\tau;\bm{x},\bm{\delta}(\cdot)))
\label{eq:levelset}
\end{equation}

\noindent where $\bm{\Phi}(t)$ defines the state trajectory under selected initial states and control trajectories. It can be seen that Eq.~\ref{eq:levelset} is equivalent to Eq.~\ref{eq:invariance} under the trim set definition Eq.~\ref{eq:trimset}. The value function $V(\bm{x},t)$ is introduced because it can be proved \cite{RN39} that $V(\bm{x},t)$ is a viscosity solution to HJB-PDE

\begin{equation}
\ppartial{V}{t}(\bm{x},t) + \min_{\tau \in [0,t]} \left\{\inf_{\bm{\delta}(\cdot)\in U(\cdot)} \ppartial{V}{\bm{x}}(\bm{x},\tau)\bm{f}(\bm{x}, \bm{\delta})\right\} = 0,
\label{eq:pde}
\end{equation}

\noindent with the initial conditions $V(\bm{x},0)=l(\bm{x})$. This proposes an optimal control method to solve for $I(T_f, K^c)$ and thus $R(T_f, K)$.

When the aircraft model is affine in the controls, i.e., $\bm{f}(\bm{x}, \bm{\delta})=\bm{b}(\bm{x})+\bm{A}(\bm{x})\bm{\delta}$, which is applicable in most cases, the optimal control selected in Eq.~\ref{eq:pde} is always one of the extreme admissible values for each control effector. The optimal control of the $i$th control effector that minimizes the Hamiltonian in Eq.~\ref{eq:pde} is given by

\begin{equation}
\delta_i^*=\left \{
\begin{matrix}[2]
\delta_{i,\max},&\ppartial{V}{\bm{x}}(\bm{x},\tau)\cdot \bm{A}_{\cdot i}(\bm{x})\leq 0\\ 
\delta_{i,\min},&\ppartial{V}{\bm{x}}(\bm{x},\tau)\cdot \bm{A}_{\cdot i}(\bm{x})>0
\end{matrix}
\right.
,
\label{eq:optu}
\end{equation}

\noindent where $\bm{A}_{\cdot i}(\bm{x})$ is the $i$th column of $\bm{A}(\bm{x})$, consisting of all control effectiveness functions for the $i$th control surface. When we evaluate this optimal control at $V(\bm{x},t)=0$, i.e., the boundary of $I(t, K^c)$ and $R(t, K)$, it gives the maximum state increment in the negative gradient direction of the boundary. This results in the largest possible increase of the reachable set boundary. So this control is also optimal in terms of enlarging the envelope. This characteristic will be useful in Sec.~\ref{sec:3.1}. Note that this framework uses open-loop commands to the system, which limits the time horizon for inherently unstable aircraft, of which flight states tends to diverge under open-loop commands.

After solving the invariant sets for both the forward and the backward systems, the envelope can be estimated by Eq.~\ref{eq:dual} and ~\ref{eq:inter}.

\subsection{Probabilistic Safe Flight Envelope}
\label{sec:2.3}

In this section, we propose an extension to the classical flight envelope with a fuzzy set, whose membership function describes the 'difficulty' to reach from and fly back to steady states for a state. Define the set of all safe flight trajectories of time horizon $T_f$ that start and end in the trim set as

\begin{equation}
S_{\bm{\Phi}}(T_f,K)=\left\{ \bm{\Phi}(t;\bm{x}(0),\bm{\delta}(\cdot)) \left\vert 
\bm{\Phi}(0) \in K, \bm{\Phi}(T_f) \in K, t\in[0,T_f]
\right. \right\}.
\end{equation}

\noindent It is noted that since aircraft can stay indefinitely in the trim set, safe flight trajectories of $T_{f}$ include all safe flight trajectories with a shorter time horizon. Then the following fuzzy set is defined as the probabilistic flight envelope.

\begin{equation}
\tilde{E}(T_f, K) = \left\{
\left(
\bm{x}\in  \mathbb{R}^n, \mu_{\tilde{E}}(\bm{x})=\dfrac{f_X(\bm{x})}{\max_{\bm{x}\in  \mathbb{R}^n} f_X(\bm{x})}
\right)
\right\},
\label{eq:prob_env}
\end{equation}

\noindent where $\mu_{\tilde{E}}(\bm{x}) \in [0,1]$ denotes the associating membership function, with $\mu_{\tilde{E}}(\bm{x})=1$ means totally safe and $\mu_{\tilde{E}}(\bm{x})=0$ means totally unsafe. $f_X(\bm{x})$ is the probability density function (PDF) of the midpoint state at $t=T_f$ for all safe flight trajectories of time horizon $2T_f$, i.e., $\bm{X}=\bm{\Phi}_X(T_f)$, $\bm{\Phi}_X$ is a random sample from $S_{\bm{\Phi}}(2T_f,K)$.

To solve for the membership function $\mu_{\tilde{E}}(\bm{x})$, the safe flight trajectories can be divided into the forward part and the backward part as

\begin{equation}
S_{\bm{\Phi},f}(2T_f,K)=\left\{ \bm{\Phi}(t) \left\vert 
\bm{\Phi} \in S_{\bm{\Phi}}(2T_f,K), t\in[0,T_f]
\right. \right\}.
\end{equation}
\begin{equation}
S_{\bm{\Phi},b}(2T_f,K)=\left\{ \bm{\Phi}(t) \left\vert 
\bm{\Phi} \in S_{\bm{\Phi}}(2T_f,K), t\in[T_f,2T_f]
\right. \right\}.
\end{equation}

\noindent Similarly, define two more PDFs $f_{\bm{X}_f}(\bm{x})$ and $f_{\bm{X}_b}(\bm{x})$ of random variables $\bm{X}_f=\bm{\Phi}_{X_f}(T_f)$, $\bm{\Phi}_{X_f}$ is a random sample from $S_{\bm{\Phi},f}(2T_f,K)$ and $\bm{X}_b=\bm{\Phi}_{X_b}(T_f)$, $\bm{\Phi}_{X_b}$ is a random sample from $S_{\bm{\Phi},b}(2T_f,K)$. Then, the membership function can be reformulated as

\begin{equation}
\mu_{\tilde{E}}(\bm{x}) = \dfrac{f_{\bm{X}_f}(\bm{x})\cdot f_{\bm{X}_b}(\bm{x})}{\max_{\bm{x}\in  \mathbb{R}^n} f_{\bm{X}_f}(\bm{x})\cdot f_{\bm{X}_b}(\bm{x})}.
\label{eq:member}
\end{equation}

It should be noted that the following fuzzy sets can be constructed as the fuzzy counterparts of the crisp forward and backward reachable set $R_f(T_f, K)$ and $R_b(T_f, K)$.

\begin{equation}
\tilde{R}_f(T_f, K) = \left\{
\left(
\bm{x}\in  \mathbb{R}^n, \mu_{\tilde{R}_f}(\bm{x})=\dfrac{f_{\bm{X}_f}(\bm{x})}{\max_{\bm{x}\in  \mathbb{R}^n} f_{\bm{X}_f}(\bm{x})}
\right)
\right\}
\end{equation}
\begin{equation}
\tilde{R}_b(T_f, K) = \left\{
\left(
\bm{x}\in  \mathbb{R}^n, \mu_{\tilde{R}_b}(\bm{x})=\dfrac{f_{\bm{X}_b}(\bm{x})}{\max_{\bm{x}\in  \mathbb{R}^n} f_{\bm{X}_b}(\bm{x})}
\right)
\right\}
\end{equation}

\begin{figure}[hbt!]
\centering
\includegraphics[width=\linewidth]{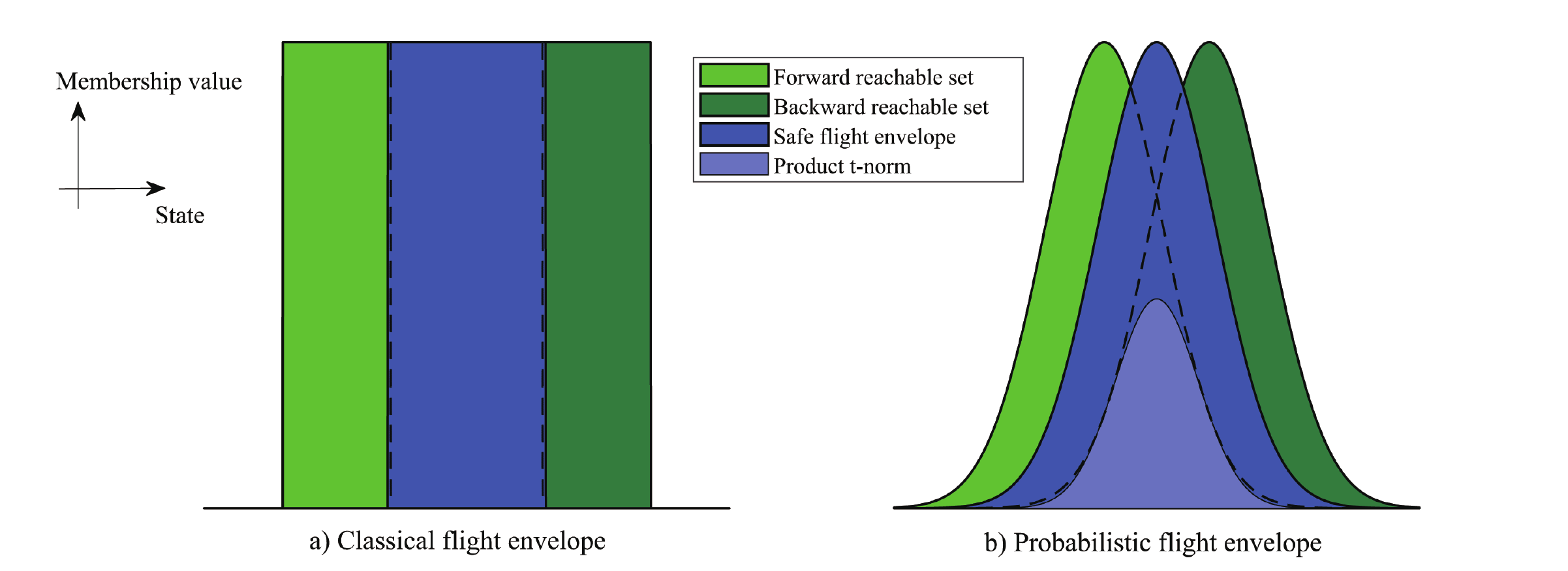}
\caption{Comparison of the classical and the probabilistic flight envelope definition for a univariate case.}
\label{fig:comp_def}
\end{figure}

Then the probabilistic flight envelope is the product t-norm of $\tilde{R}_f(T_f, K)$ and $\tilde{R}_b(T_f, K)$ scaled by a factor such that the center of the envelope has a membership value of 1. The difference between the crisp set and the fuzzy set definition of the safe flight envelope is illustrated in Fig.~\ref{fig:comp_def} for a univariate case. 

\section{Envelope Estimation by Monte Carlo Simulation}
\label{sec:3}

MC simulations are applied to obtain a computationally-feasible estimation of the probabilistic flight envelope defined by Eq.~\ref{eq:prob_env} with high dimensions. MC simulation is a computational algorithm that uses random sampling to obtain a numerical estimation of probability distributions \cite{RN42}, which has not been applied to envelope estimation. The closest application is the workspace determination in robotics \cite{RN41}.

\begin{figure}[hbt!]
\centering
\includegraphics[width=3.25in]{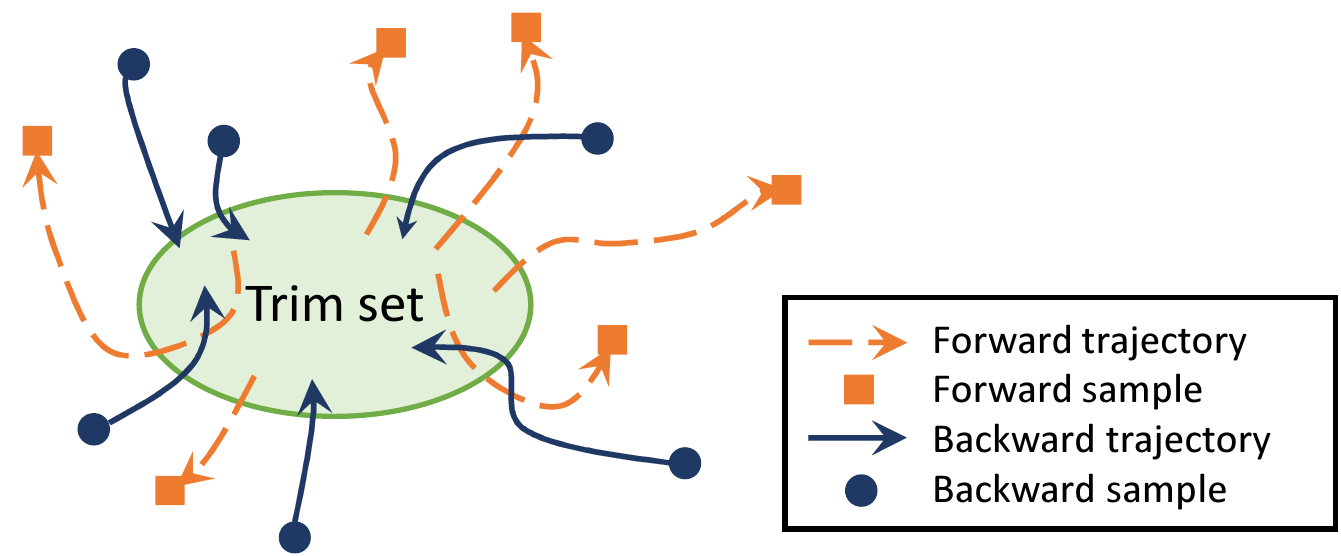}
\caption{Illustration of the sampling process in MC simulations.}
\label{fig:mc_sim}
\end{figure}

\begin{figure}[hbt!]
\centering
\includegraphics[width=3.25in]{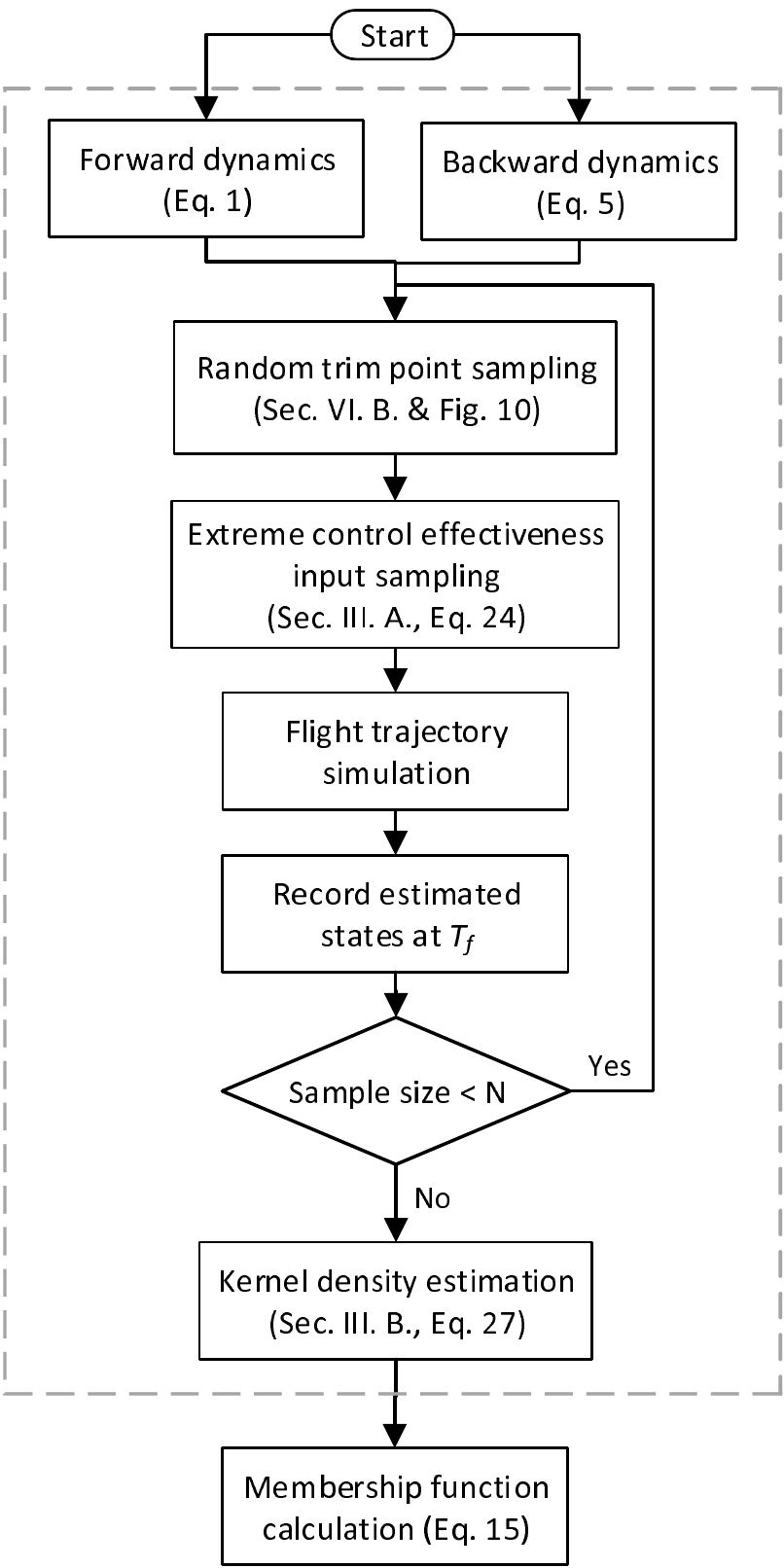}
\caption{Flow chart of envelope estimation with MC simulation.}
\label{fig:mc_flow}
\end{figure}

To estimate the envelope, two MC simulation routines are performed to obtain the probability distributions of $\bm{X}_f$ and $\bm{X}_b$ respectively. First, a random point in the trim set is selected as the initial state. Then, pseudo-random open-loop control inputs are selected, which should be representative to explore the boundary of the flight envelope. The selection method is discussed in Sec.~\ref{sec:3.1}. With the initial state and control inputs specified, a flight trajectory of time horizon $T_f$ can be simulated by solving the forward dynamics (Eq.~\ref{eq:forward}) or the backward dynamics (Eq.~\ref{eq:backward}) under desired flight conditions. The states at time $T_f$, of which the envelope is to be estimated are stored as one sample. This sampling process is illustrated in Fig.~\ref{fig:mc_sim}. With a large size of samples generated, the probability distributions of $\bm{X}_f$ and $\bm{X}_b$ can be estimated by kernel density estimation, details of which are discussed in Sec.~\ref{sec:3.2}. Finally, the membership function of the probabilistic flight envelope is calculated by Eq.~\ref{eq:member}. A flowchart of the whole process is shown in Fig.~\ref{fig:mc_flow}.

\subsection{Extreme Control Effectiveness Method}
\label{sec:3.1}

For a full degree-of-freedom aircraft model, only the dynamics of rotational and translational velocities are directly dependent on the control surface deflections via aerodynamic forces and moments. Define the effective states in envelope estimation as 

\begin{equation}
\bm{x}_e = [u\ v\ w\ p\ q\ r]^T.
\end{equation}

\noindent The dynamics of $\bm{x}_e$ can be expressed in the incremental form \cite{RN25} as

\begin{equation}
\bm{f}(\bm{x}_e, \bm{\delta}) \approx \dot{\bm{x}}_{e,0} + \tilde{\bm{J}}^{-1} \bm{J}_d\Delta\bm{\delta},
\label{eq:incdyn}
\end{equation}

\noindent where $\dot{\bm{x}}_{e,0}$ is the state derivatives at the current time step, $\Delta\bm{\delta}$ is the incremental deflections of all active control surfaces and the incremental thrust,

\begin{equation}
\tilde{\bm{J}}^{-1}=
\begin{bmatrix}
\mathrm{diag}\left(\frac{1}{m},\frac{1}{m},\frac{1}{m}\right) & \bm{0}\\ 
\bm{0} & \bm{J}^{-1}
\end{bmatrix},
\end{equation}

\begin{equation}
\bm{J}_d = 
\begin{bmatrix}
\bm{J}_d^F\\
\bm{J}_d^M
\end{bmatrix}
=
\left[\left(\frac{\partial F_x}{\partial \bm{\delta}}\right)^T \left(\frac{\partial F_y}{\partial \bm{\delta}}\right)^T \left(\frac{\partial F_z}{\partial \bm{\delta}}\right)^T \left(\frac{\partial M_x}{\partial \bm{\delta}}\right)^T \left(\frac{\partial M_y}{\partial \bm{\delta}}\right)^T \left(\frac{\partial M_z}{\partial \bm{\delta}}\right)^T\right]^T.
\end{equation}

\noindent Note that the incremental setup is only necessary to generalize complex systems. Eq.~\ref{eq:incdyn} can be replaced by other non-incremental affine dynamics. Thus, according to Eq.~\ref{eq:optu}, we have

\begin{equation}
\ppartial{V}{\bm{x}}(\bm{x},\tau)\cdot \bm{A}_{\cdot i}(\bm{x})=\ppartial{V}{\bm{x}_e}(\bm{x}, \tau)\tilde{\bm{J}}^{-1}\bm{J}_{d,\cdot i},
\end{equation}

\begin{equation}
\Delta \delta_i^*=\left \{
\begin{matrix}[2]
\Delta \delta_{i,\max},&\dfrac{\partial V}{\partial \bm{x}_e}(\bm{x}, \tau)\tilde{\bm{J}}^{-1}\bm{J}_{d,\cdot i}<0\\ 
\Delta \delta_{i,\min},&\dfrac{\partial V}{\partial \bm{x}_e}(\bm{x}, \tau)\tilde{\bm{J}}^{-1}\bm{J}_{d,\cdot i}>0
\end{matrix}
\right. .
\label{eq:optu2}
\end{equation}

When applying MC simulations, the only unknown term in Eq.~\ref{eq:optu2} is the gradient of the level set function $\partial V/\partial \bm{x}_e$,  which determines the exploration direction that enlarges the envelope the most. The optimal control thus optimizes the increment in this direction. Instead of solving the HJB-PDE globally for the optimal exploration strategy, a random exploration strategy along the trajectory is applied by replacing this variable with a random vector $\bm{W}$ such that each direction has the same probability of being selected. This can be done with an identical distribution symmetric around zero (e.g. $N(0,1)$) for each element $W_j$. This random vector is sampled at each time step. So, the following strategy of optimal control selection is proposed for the extreme control effectiveness method.

\begin{equation}
\Delta \delta_i^*=\left \{
\begin{matrix}
\Delta \delta_{i,\max},&\bm{W}\tilde{\bm{J}}^{-1}\bm{J}_{d,\cdot i}<0\\ 
\Delta \delta_{i,\min},&\bm{W}\tilde{\bm{J}}^{-1}\bm{J}_{d,\cdot i}>0
\end{matrix}
\right. , W_j \thicksim N(0,1),
\label{eq:optu3}
\end{equation}

\noindent where $j\in\left\{u,v,w,p,q,r\right\}$. In this way, the population to sample the control inputs reduces significantly from all admissible control inputs $U$ to limited combinations of extreme control input increments. This reduced population, however, still includes all possible optimal control selections. So the convex hull of the optimized samples still converges to the envelope boundary defined by $V(x,T_f)=0$.

This method is named extreme control effectiveness method, since Eq.~\ref{eq:optu3}, in essence, aims to optimize a weighted sum of control effectiveness in three translational and three rotational directions. So the control input samples generated with this method explore the flight envelope in a particular direction of the vector $\bm{W}=[W_u\ W_v\ W_w\ W_p\ W_q\ W_r]$ with extreme control effectiveness. This method is pseudo-random because it samples exploration directions rather than the actual control inputs by always assigning corresponding extreme control inputs.

\subsection{Kernel Density Estimation}
\label{sec:3.2}
Density estimation constructs estimators of the probability density functions $f_{\bm{X}_f}(\bm{x})$ and $f_{\bm{X}_b}(\bm{x})$ based on samples of $\bm{X}_f$ and $\bm{X}_b$. Since no prior knowledge is available about the structure of the envelope, the famous non-parametric method kernel density estimation \cite{RN12} is applied to generate a continuous estimator of the probabilitic flight envelope.

To estimate the density of a $d$-dimensional random vector 

\begin{equation}
\bm{X} = (X_1,X_2,...,X_d)^T
\end{equation}

\noindent by $N$ samples 

\begin{equation}
\bm{y}_i = (y_{i1},y_{i2},...,y_{id})^T, i = 1,2,...,N,
\end{equation}

\noindent the multivariate kernel density estimator is given in Ref.~\cite{RN12} as

\begin{equation}
\hat{f}(\bm{x}) = \frac{1}{N}\sum_{i=1}^N K_H(\bm{x} - \bm{y}_i) = \frac{1}{Nh_1h_2...h_d}\sum_{i=1}^N\prod_{j=1}^d k\left(\frac{x_j-y_{ij}}{h_j}\right),
\label{eq:kernel1}
\end{equation}

\noindent where $K_H(\cdot)$ is the kernel, which describes the estimated distribution with one sample at the origin. $k(\cdot)$ is the normalized kernel function for each dimension satisfying

\begin{equation}
\int_{-\infty}^{\infty}k(\zeta)\mathrm{d}\zeta = 1.
\end{equation}

\noindent Usually, a Gaussian kernel function 

\begin{equation}
k(\zeta) = \frac{1}{\sqrt{2\pi}}e^{-\frac{1}{2}\zeta^2}
\end{equation}

\noindent is used. The bandwidth $h_j$ is comparable to the bin size in the histogram, which normalizes the kernel by characterizing  the range affected by one sample in the estimator. The bandwidth can be selected by Silverman's rule of thumb \cite{RN12}

\begin{equation}
h_j = \sigma_j\left[\frac{4}{(d+2)N}\right]^{1/(d+4)},
\label{eq:kernel2}
\end{equation}

\noindent where $\sigma_j$ is the standard deviation of the $j$th element in the sample.

\subsection{Advantages and Limitations}

In addition to tractability, this MC-simulation-based envelope estimation method demonstrates additional advantages over conventional reachability analysis. First, the conservativeness of the estimated envelope is guaranteed, since sample points are associated with particular flight trajectories and control inputs for reconstruction, whereas for the level set method, only the indirect evolution of the level set function is concerned. Second, the estimated envelope is probabilistic in the form of fuzzy sets, which gives additional information about how safe a certain point within the envelope is. This is especially useful in FEP system design. The aggressiveness of FEP can be altered by specifying different $\alpha$-cuts of the fuzzy set (i.e., different thresholds of the membership function), depending on applications. Furthermore, the approaching of the envelope boundary is detectable in advance by the gradient of the membership function as shown by the predictive probabilistic FEP law proposed in Sec.~\ref{sec:4.2}. Finally, this method provides great flexibility to design desired envelope databases. The exact same routine can be applied to any set of states to be protected. Simulations can be augmented straightforwardly with additional external and internal stochastic components, since the estimation already is stochastic.

However, it should be noted that this method trades optimality for feasibility. Despite the fact that a conservative estimation is always available, the required sample size to achieve the same level of optimality still suffers the curse of dimensionality. Nevertheless, the base of the exponential complexity for the MC-simulation-based method is usually smaller than that for the level set method. It was shown in Ref.~\cite{RN81} that the required sample size to maintain the same level of the mean integrated squared error (MISE) in kernel density estimation is

\begin{equation}
N_{\mathrm{req}} = \mathcal{O}\left(\mathrm{MISE}^{-\frac{4+d}{4}}\right).
\end{equation}

\noindent For comparison, the level set method has a time complexity of $\mathcal{O}\left(N_g^{d+1}\right)$, where $N_g$ is the number of computational grid points in each dimension \cite{Zhang_2019}. So the envelope estimation with MC simulation has a speedup factor of

\begin{equation}
\eta = \left(N_g \cdot \sqrt[4]{\mathrm{MISE}}\right)^d
\end{equation}

\noindent over the level set method. For example, when $N_g=20$ and $\mathrm{MISE}=\num{1e-3}$, the speedup factor $\eta \approx 3.56^d$ for a $d$-dimensional problem.

\section{FEP System Design with Probabilistic Envelope}
\label{sec:4}

In general, online FEP systems with probabilistic envelope estimated offline consist of two parts: 1) an envelope database that generates an envelope metric at the current state in real-time as $\bm{M}_{\mathrm{env}}=\bm{M}_{\mathrm{env}}(\bm{x})$ and 2) FEP laws that modify the references of protected states based on the envelope metric as $\bm{x}_{\mathrm{fep}}=\bm{x}_{\mathrm{fep}}\left(\bm{x}_{\mathrm{ref}},\bm{M}_{\mathrm{env}}\right)$. The estimated probabilistic envelope can be protected by both conventional command limiting and novel probabilistic command modification, both of which are introduced as follows.

\subsection{State-Constraint-Based FEP}
\label{sec:4.1}

The state-constraint-based FEP starts by converting the probabilistic envelope to a classical envelope with absolute boundaries. The threshold of the membership function is parametrized by $k_0$ as

\begin{equation}
\mu_{{\tilde{E}},0} = e^{-\frac{1}{2}k_0^2}.
\label{eq:binary}
\end{equation}

\noindent The $k_0$ value can be specified depending on the level of allowed LOC risk. The envelope metric is then constructed as a matrix of maximum and minimum constraints of protected states as

\begin{equation}
\bm{M}_{\mathrm{env}}(\bm{x})=
\left[\bm{x}_{\min}(\bm{x})\  
\bm{x}_{\max}(\bm{x})\right].
\label{eq:menv1}
\end{equation}

\begin{figure}[hbt!]
\centering
\includegraphics[width=\linewidth]{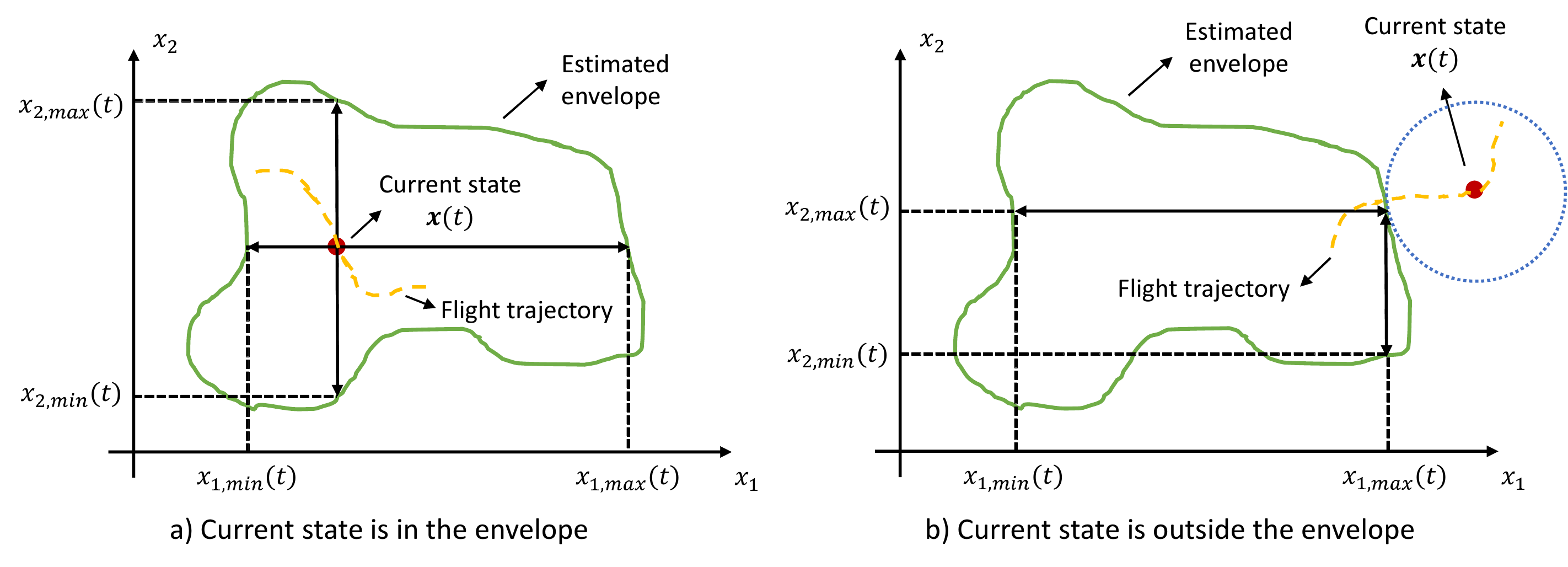}\\
\caption{Definition of dynamic state constraints for a bivariate case.}
\label{fig:constraint}
\end{figure}

\noindent These state constraints are defined as follows. When the current state is in the estimated envelope, the constraints are defined as the maximum and minimum values for each state to stay within the envelope with the other states unchanged. When the current state is outside the estimated envelope, the constraints are defined the same as the closest point that is within the envelope. This definition is illustrated in Figure~\ref{fig:constraint} for a bivariate case.

The FEP law for the state-constraint-based method is direct command limiting. The envelope-protected reference is a saturation function of the original reference with the upper and lower limits defined in Eq.~\ref{eq:menv1} as shown in the following equation.

\begin{equation}
\bm{x}_{\mathrm{fep}} = \max\left[\bm{x}_{\min}, \min\left(\bm{x}_{\max}, \bm{x}_{\mathrm{ref}}\right)\right],
\label{eq:fep1}
\end{equation}

\subsection{Probabilistic FEP}
\label{sec:4.2}

For probabilistic FEP, the envelope metric stored in the envelope database is constructed as a monotonic mapping from the membership value $\mu_{\tilde{E}}(\bm{x})$ empirically with

\begin{equation}
M_{\mathrm{env}}(\bm{x}) = \ln \left(\max \left(\mu_{\tilde{E}}(\bm{x}), \epsilon\right)\right),
\end{equation}

\noindent where $\epsilon$ is a small constant. This metric describes how much the current state needs to be protected. This is determined by two factors: 1) the magnitude of $M_{\mathrm{env}}(\bm{x})$, which describes how dangerous the current state is, and 2) the gradient of $M_{\mathrm{env}}(\bm{x})$, which describes how effective the protection will be. Thus, the probabilistic FEP law is proposed as 

\begin{equation}
\bm{x}_{\mathrm{fep}} = \bm{x}_{\mathrm{ref}} + \bm{\chi}(\bm{x}),
\label{eq:fep2}
\end{equation}

\noindent with

\begin{equation}
\bm{\chi}(\bm{x}) = 
\left\{\begin{matrix}
-\left(M_{\mathrm{env}}(\bm{x})-M_0\right)\bm{K}_{\mathrm{fep}}\ \bm{J}^T_{\mathrm{env}}(\bm{x}),&M_{\mathrm{env}}(\bm{x})\leq M_0\\ 
0,&M_{\mathrm{env}}(\bm{x})>M_0
\end{matrix}\right.
,
\end{equation}

\noindent where $M_0$ is the threshold to activate envelope protection, $\bm{K}_{\mathrm{fep}}$ is a diagonal gain matrix, $\bm{J}_{\mathrm{env}} = \partial M_{\mathrm{env}} / \partial \bm{x}$ is the gradient of the envelope metric.

\begin{figure}[hbt!]
\centering
\includegraphics[width=\linewidth]{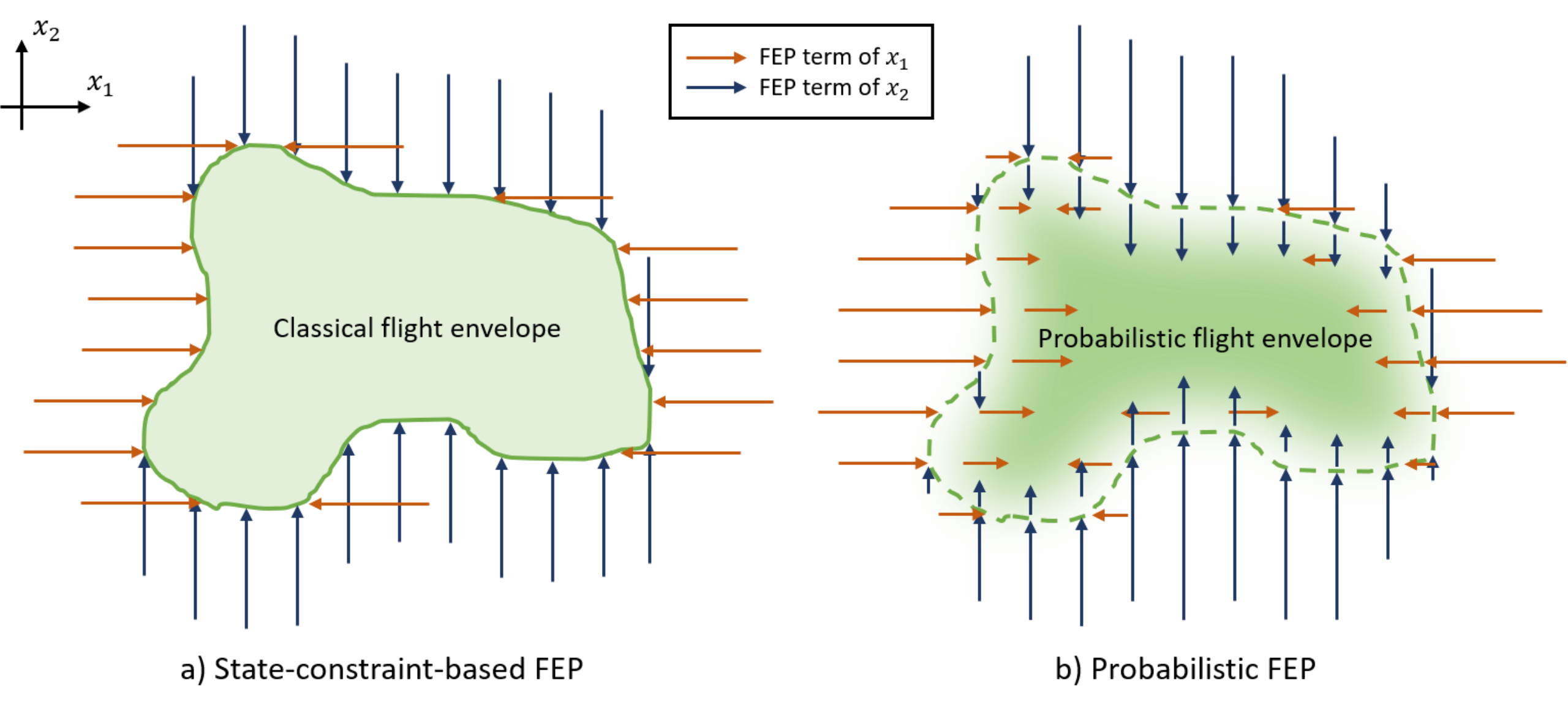}
\caption{Comparison of FEP strategies for state-constraint-based and probabilistic methods.}
\label{fig:comp_fep}
\end{figure}

Comparing both FEP methods, the advantages of the probabilistic FEP are that 1) the aggressiveness of FEP can be altered online by tuning the threshold $M_0$ and the gain vector $\bm{K}_{\mathrm{fep}}$ without re-calculating the whole envelope database, 2) the modification to the command is predictive, which starts before reaching the envelope boundary, and 3) the boundedness of the modification term is guaranteed, which completes the boundedness proof of PCH as will be shown in Sec.~\ref{sec:5}. Conversely, the state-constraint-based FEP guarantees to constrain the command within a fixed region in the state space, so its behavior is more predictable. A comparison of the applied FEP strategies for both methods is illustrated in Fig.~\ref{fig:comp_fep} for a bivariate case. 

\section{Multi-Loop NDI Controllers with Command Modification}
\label{sec:5}

Both types of FEP systems interact with the controller by modifying the command. This command modification is trivial for single-loop controllers as the command can be arbitrarily selected. However, the FCS usually utilizes the time-scale separation principle to simplify the controller design \cite{Menon_1997} and is thus multi-loop. This section highlights the inner-loop command modification problem with multi-Loop NDI controllers.

\subsection{The Concept of NDI and Incremental NDI}

To tackle problems of complex gain scheduling and low robustness to model inaccuracy and failure observed in conventional linear controllers, NDI was proposed to inherently include nonlinearity in control laws by inverting the flight dynamics \cite{slotine1991applied}. Consider an $n$-loop affine system

\begin{equation}
\dot{\bm{x}}_i = \bm{b}_i(\bm{x}) + \bm{A}_i(\bm{x})\bm{x}_{i+1},i=1,2,...,n,
\label{eq:sys}
\end{equation}

\noindent where $\bm{x}_1,\bm{x}_2,...,\bm{x}_n$ are time-scale-separated states, $\bm{x}_{n+1}=\bm{\delta}$ is the control inputs. The multi-loop NDI controller of the system consists of two parts for each loop: 1) a linear PID controller that generates the virtual control inputs as

\begin{equation}
\bm{\nu}_i(\bm{x})=\dot{\bm{x}}_i=\bm{K}_{\mathrm{p},i}\bm{e}_i+\bm{K}_{\mathrm{i},i}\int_0^t\bm{e}_i\mathrm{d}t+\bm{K}_{\mathrm{d},i}\dfrac{\mathrm{d}\bm{e}_i}{\mathrm{d}t},
\end{equation}

\noindent where $\bm{e}_i=\bm{x}_{i,\mathrm{com}}-\bm{x}_i$ is the tracking error, and 2) input-output linearization by inverting the dynamics

\begin{equation}
\bm{x}_{i+1,\mathrm{com}} =  \bm{A}_i^{-1}(\bm{x})\left(\bm{\nu}_i(\bm{x}) - \bm{b}_i(\bm{x})\right),
\label{eq:io}
\end{equation}

\noindent that maps the virtual control of the $i$th loop to the command of the $(i+1)$th loop or the final control inputs.

In order to handle model inaccuracy, incremental NDI was developed in Ref.~\cite{RN61, RN62} by applying NDI to the incremental form of the dynamic equations, such that only the control effectiveness part of the model is relevant in the controller design to robustify the controller. The incremental approach applies first-order Taylor series expansion and time-scale separation to the system (Eq.~\ref{eq:sys}) at the current states and inputs as

\begin{equation}
\dot{\bm{x}}_i \approx \dot{\bm{x}}_{i,0} + \bm{A}_i(\bm{x}_0)\Delta\bm{x}_{i+1},i=1,2,...,n,
\label{eq:sysi}
\end{equation}
\begin{equation}
\bm{x}_{i+1,\mathrm{com}} =  \bm{A}_i^{-1}(\bm{x}_0)\left(\bm{\nu}_i(\bm{x}) - \dot{\bm{x}}_{i,0}\right) + \bm{x}_{i+1,0},i=1,2,...,n.
\label{eq:ioi}
\end{equation}

It should be noted that the incremental approach proposes additional requirements on the control system \cite{RN61}: 1) the controller should have access to measurements of state derivatives (especially derivatives of angular rates), and 2) the system should have both fast sampling rates and fast control actions. Therefore, a combined approach is adopted with the incremental approach for dynamic loops and the ordinary approach for kinematic loops whose equations of motion are universal with no uncertainties.

\subsection{Multi-Loop NDI with Pseudo Control Hedging}

In the general framework of multi-loop NDI, outer loops assume inner-loop tracking with high bandwidth. However, inner-loop tracking here is sacrificed by command modification from FEP. Additional feedback to outer loops is thus needed to avoid continuous activation of FEP command modification. Unlike in command-filtered backstepping \cite{RN64, RN65} where asymptotic tracking is always desired, this feedback is more analogous to the input saturation avoidance problem in single-loop configurations to achieve bounded-input, bounded-output (BIBO) stability. One such technique is PCH. PCH was originally proposed to avoid the effect of input saturation on system identification \cite{RN68}. The concept was later expanded to general flight control systems to avoid and compensate for actuator saturation of control surfaces \cite{RN66, RN69}. Ref.~\cite{RN68} proved the boundedness of PCH with bounded command modification. The concept of PCH is generalized to multi-loop as follows.

\begin{figure}[hbt!]
	 \centering
	 \includegraphics[width=\linewidth]{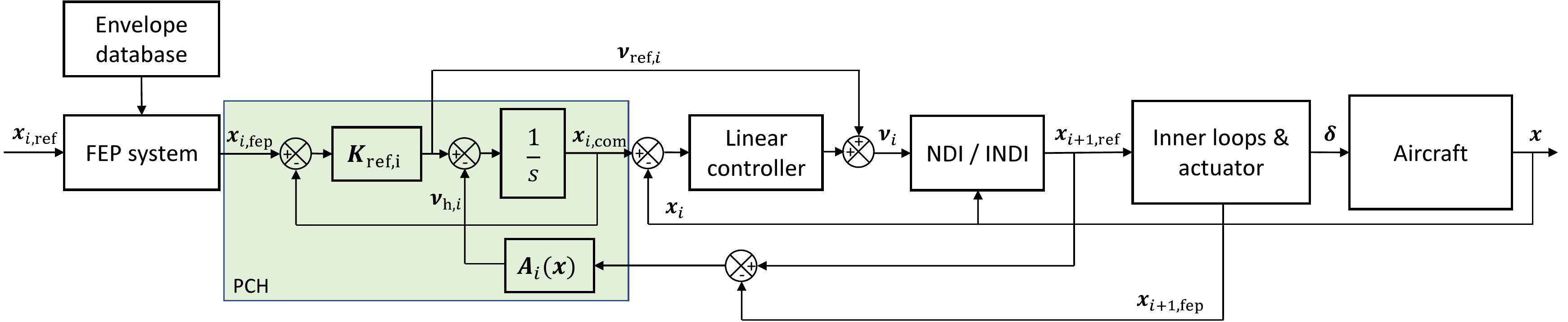}
	 \caption{Block diagram of multi-loop NDI controllers with FEP and PCH.} 
	 \label{fig:pch}
\end{figure}

Consider an affine system as Eq.~\ref{eq:sys}. The block diagram of multi-loop NDI controllers with FEP and PCH is shown in Fig.~\ref{fig:pch}. As shown in Fig.~\ref{fig:pch}, the original command generated by outer-loop NDI is now known as the reference signal $\bm{x}_{i,\mathrm{ref}}$. This signal is protected by state-constraint-based or probabilistic FEP system with Eq.~\ref{eq:fep1} or \ref{eq:fep2}, and becomes the envelope-protected signal $\bm{x}_{i,\mathrm{fep}}$. Then, the envelope-protected signal is compensated by PCH based on the amount of inner-loop protection to be the final command $\bm{x}_{i,\mathrm{com}}$ to the inner loop. The following reference model is used for PCH.

\begin{equation}
	\dot{\bm{x}}_{i,\mathrm{com}} = \bm{K}_{\mathrm{ref},i} (\bm{x}_{i,\mathrm{fep}} - \bm{x}_{i,\mathrm{com}}) - \bm{\nu}_{\mathrm{h},i} = \bm{\nu}_{\mathrm{ref},i} - \bm{\nu}_{\mathrm{h},i},
\end{equation}

\noindent where 

\begin{equation}
	\bm{\nu}_{\mathrm{h},i} = \bm{A}_i(\bm{x})\left(\bm{x}_{i+1,\mathrm{ref}} - \bm{x}_{i+1,\mathrm{fep}}\right)
\end{equation}

\noindent is the hedged virtual control due to inner-loop FEP, $\bm{K}_{\mathrm{ref},i} $ is the reference model gain, which is a design parameter usually selected the same as the linear controller gain. In addition, $\bm{\nu}_{\mathrm{ref},i}$ is fed forward to $\bm{\nu}_i$ to retain the tracking performance.

\section{Implementation}
\label{sec:7}

\subsection{The ICE Aircraft Model}
\label{sec:6.2}

\begin{figure}[hbt!]
	 \centering
	 \includegraphics[width=3.25in]{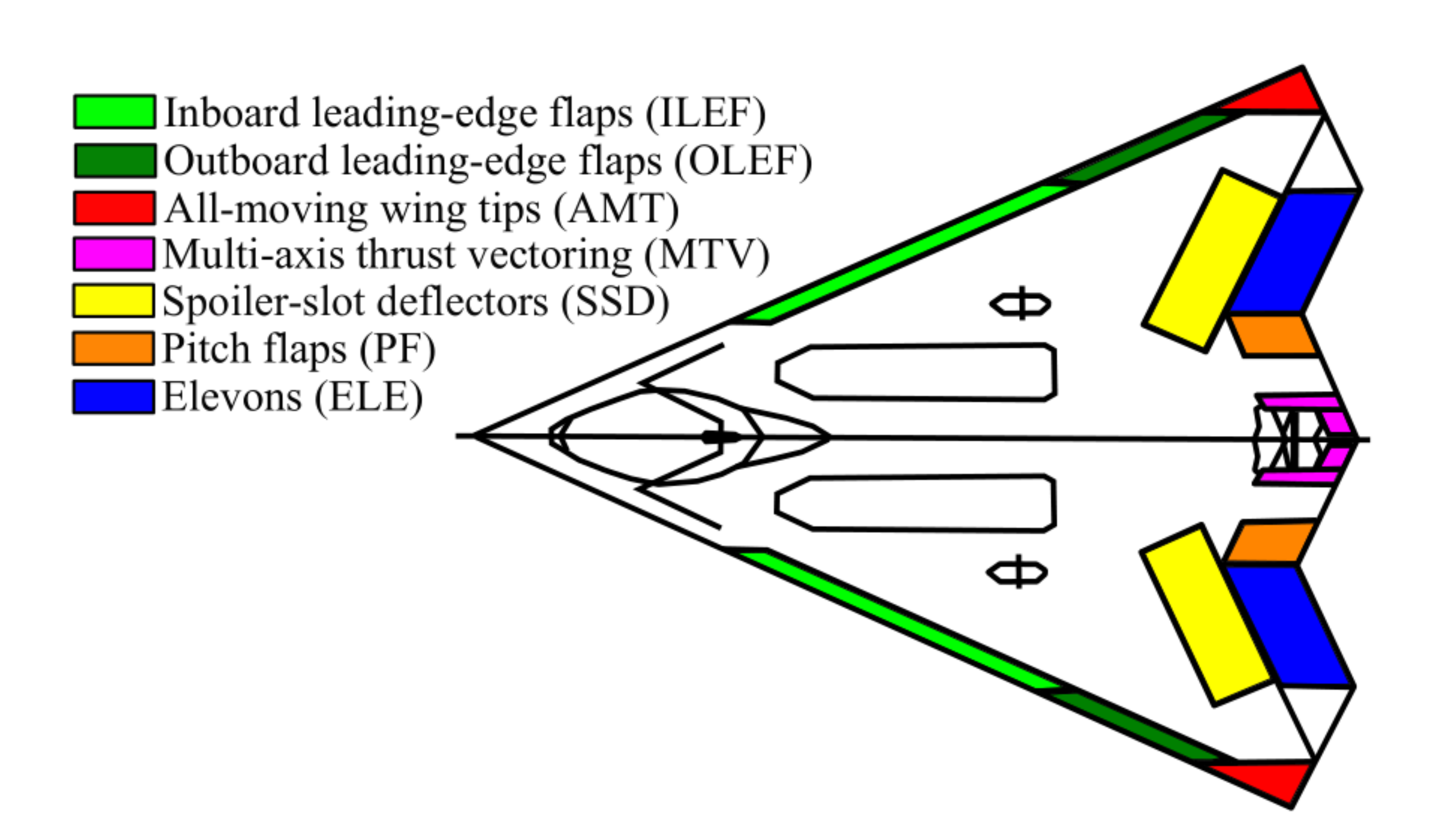}
	 \caption{Control effector configuration of the ICE aircraft \protect\cite{RN25}.} 
	 \label{fig:ice}
\end{figure}

The proposed probabilistic envelope estimation and protection framework was implemented on a simulation model of Lockheed Martin’s ICE aircraft. The ICE project aims to explore novel control surfaces for high-performance tailless fighter aircraft \cite{RN46}. This paper focuses on a land-based configuration of the ICE aircraft (ICE 101-TV). The control suite on the ICE aircraft consists of 13 control effectors of six types, namely inboard and outboard leading-edge flaps (LEF), all-moving wing tips (AMT), fluidic multi-axis thrust vectoring (MATV), spoiler-slot deflectors (SSD), pitch flaps (PF), and elevons. The configuration of these control effectors is illustrated in Figure~\ref{fig:ice}. Due to page limits, readers are guided to Ref.~\cite{RN25,RN80} for details of the aircraft and its model.

Depending on their characteristics, different positional and no-load rate limits are specified for the control effectors. The upper and lower boundaries of the control surface deflections are then

\begin{equation}
\left\{
\begin{aligned}
\bm{\delta}_{\mathrm{u}}(t) &= \min\left(\bm{\delta}_{\max},\bm{\delta}_0+\dot{\bm{\delta}}_{\max}\Delta t\right)\\
\bm{\delta}_{\mathrm{l}}(t) &= \max\left(\bm{\delta}_{\min},\bm{\delta}_0-\dot{\bm{\delta}}_{\min}\Delta t\right)
\end{aligned}
\right.
,
\end{equation}

\noindent where $\bm{\delta}_{\max}, \bm{\delta}_{\min}$ are the position limits, $\dot{\bm{\delta}}_{\max}, \dot{\bm{\delta}}_{\min}$ are the rate limits. These boundaries define the set of admissible input functions $U(t)$.

A high-fidelity aerodynamic model was released based on wind tunnel tests from Lockheed Martin. The core part of the model describes the aerodynamic coefficients of forces and moments with respect to states and control surface deflections. These coefficients are expressed by summations of terms which model the contribution of the base airframe, aerodynamic control surfaces, and interactions between control surfaces as

\begin{equation}
C_i = \sum_{j}C_{ij}\left(\bm{\delta},M,\alpha,\beta,p,q,r\right)
\end{equation}

\noindent where $i\in\left\{F_x,F_y,F_z,M_x,M_y,M_z\right\}$. A total of 108 nonlinear terms $C_{ij}$ are stored in look-up tables (LUT) and interpolated by cubic spline interpolation. Details of the aerodynamic database can be found in Ref.~\cite{RN80}. Three predefined mass configurations can be used for light-load, nominal, and heavy-load cases. The above databases and models are embedded in Simulink that solves the equations of motion in real-time at 100Hz.

Previous research parametrized the aerodynamic database with multivariate splines \cite{RN26} and developed an incremental nonlinear control allocation (INCA) scheme which effectively allocates the desired aerodynamic forces and moments to control effectors \cite{RN25}. The multivariate spline model identified each of the nonlinear terms $C_{ij}$ with spline functions of polynomial degree 3 to 5 and 0th order continuity to obtain a well-defined first order spatial derivative over the entire flight envelope. Thus, the control effectiveness matrix $\bm{J}_d$ can be updated efficiently in real-time by summing up the gradient of each spline model analytically for each aerodynamic control effector.

\subsection{Trim Set}

As the starting point of envelope estimation, the trim set of the ICE aircraft needs to be specified as the apriori safe set $K$. Trimming aims to find the equilibrium points of the system by formulating a constrained optimization problem that constrains state derivatives to zero. 

This paper considers steady-state, straight, and level flights at altitudes between 10000 ft and 30000 ft and Mach numbers between 0.4 and 1.2 as the trim set of the ICE aircraft. Individual trim points are optimized on an altitude-Mach grid within this range. The trim points are then interpolated by cubic splines to form a continuous trim set. Note that all points in the 2-D trim set are relevant as boundary points in the high-dimensional estimation space. Only AMTs, elevons, and pitch thrust vectoring are used to trim the aircraft for efficiency. Note that AMTs and elevons deflect symmetrically on a level flight. Therefore, the following variables are optimized

\begin{equation}
\bm{x}_{\mathrm{trim}} = [T\ d_{\mathrm{amt}}\ d_{\mathrm{ele}}\ d_{\mathrm{ptv}}\ \alpha]^T,
\end{equation}

\noindent with the following constraints

\begin{equation}
\dot{p}=\dot{q}=\dot{r}=\dot{u}=\dot{v}=\dot{w}=\phi=\theta-\alpha=0.
\end{equation}

\noindent The objective function is formulated as

\begin{equation}
J = J_1 + k_{\mathrm{trim}} \cdot J_2,
\end{equation}

\noindent where $k_{\mathrm{trim}}$ is a weighting factor,

\begin{equation}
J_1 = \left( \frac{T}{T_{\mathrm{max}}}\right) ^2, J_2 = d_{\mathrm{amt}} \mathrm{[rad]} ^2 + d_{\mathrm{ele}} \mathrm{[rad]} ^2 + d_{\mathrm{ptv}} \mathrm{[rad]} ^2
\end{equation}

\noindent are the objectives of minimum thrust and minimum control effort respectively.

\begin{figure}[hbt!]
	 \centering
	 \includegraphics[width=3.25in]{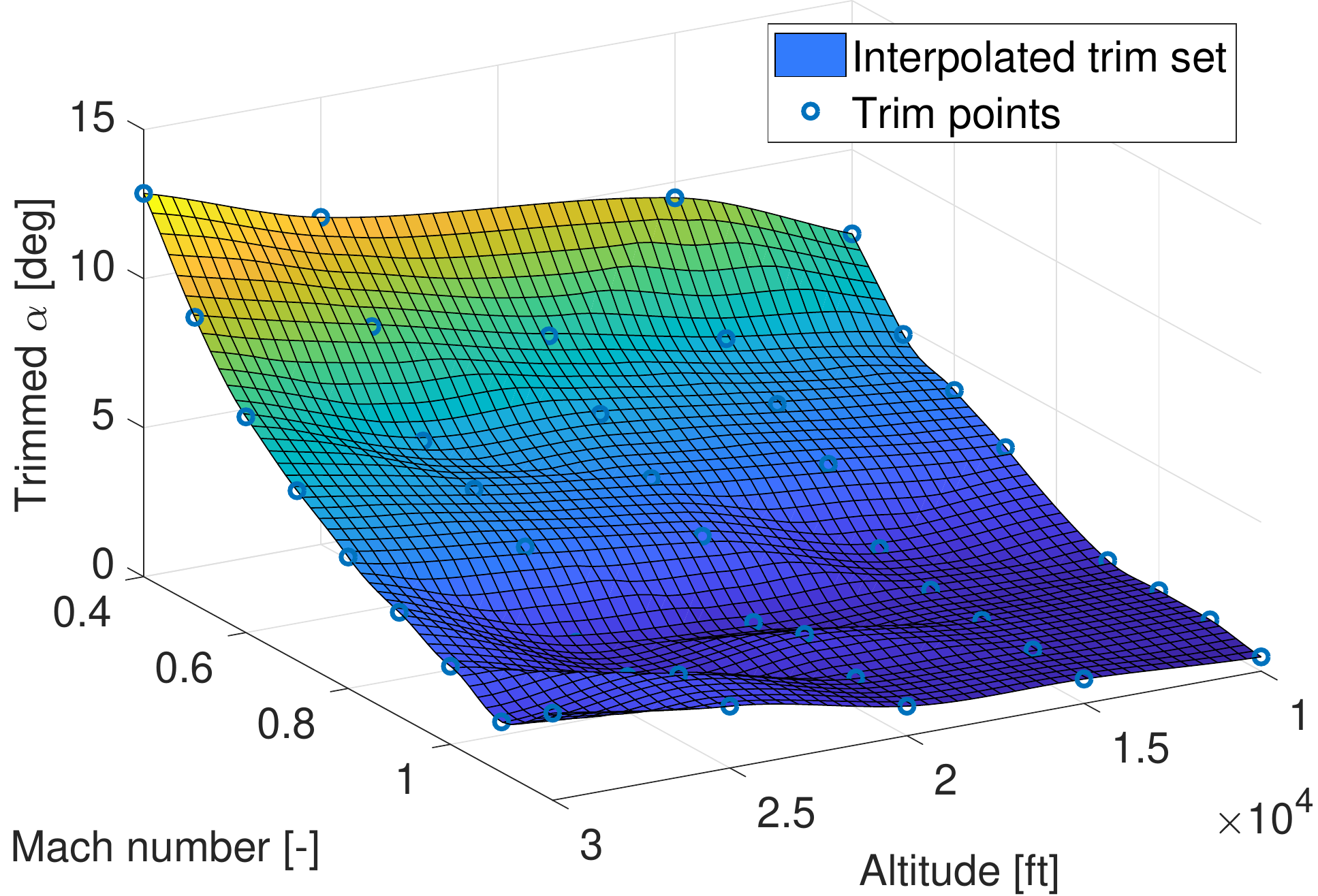}
	 \caption{Trimmed AoA at different altitudes and speeds with nominal mass configuration.} 
	 \label{fig:trim}
\end{figure}

This constrained optimization problem is solved by a sequential quadratic programming solver. The trimmed AoA at the nominal mass configuration is shown in Fig.~\ref{fig:trim}. This process is repeated for different mass configurations and the case without thrust vectoring to compare envelope estimation results in Sec.~\ref{sec:8}. When thrust vectoring is not activated, PFs are used to trim the aircraft.

\subsection{State Definition}

This paper considers seven states in the FEP of the ICE aircraft, including altitude $h$, ground speed $V_g$, aerodynamic angles $\alpha, \beta$, and angular rates $p,q,r$, which can well characterize the agility of the aircraft. It should be noted that roll angle $\phi$ and pitch angle $\theta$ are not included. This is because high-performance fighter aircraft like the ICE aircraft should be able to steer to virtually any attitude. Instead, angular rates are protected to limit specific dangerous maneuvers. In this way, a seven-dimensional envelope database was established. To avoid instability, the time horizon is limited to $T_f \leq$ 1.5s. This time horizon is smaller than the characteristic time scales of altitude $h$ and ground speed $V_g$. Thus, the altitude and the speed are not protected in the applied FEP systems, but only used as parameters to obtain the envelope of the other five states.

\subsection{Controller Structure}

The controller implemented in this paper is a two-loop controller, which consists of an outer loop of aerodynamic angles $\phi, \alpha, \beta$ with ordinary NDI and an inner loop of angular rates $p,q,r$ with incremental NDI. The outer loop is selected to test high AoA maneuvers for which the ICE aircraft is optimized. So, we have 

\begin{equation}
\bm{x}_1=[\phi\ \alpha\ \beta]^T, \bm{x}_2=[p\ q\ r]^T, \bm{x}_3=\bm{\delta}.
\end{equation}

The dynamics of the outer loop are

\begin{equation}
	\begin{aligned}
    	\dot{\bm{x}}_1
    &
        =
        \dfrac{\mathrm{d}}{\mathrm{d}t}
        \begin{bmatrix}
        \phi\\
        \alpha\\ 
        \beta\\
        \end{bmatrix}
        =
        \begin{bmatrix}
        0 \\
        \dfrac{1}{u^2+w^2}\left(u\tilde{A}_z-w\tilde{A}_x\right) \\[6pt]
        \dfrac{1}{\sqrt{u^2+w^2}}\left[\dfrac{-uv}{V_g^2}\tilde{A}_x+(1-\dfrac{v^2}{V_g^2})\tilde{A}_y-\dfrac{vw}{V_g^2}\tilde{A}_z\right]
        \end{bmatrix}
        +
        \begin{bmatrix}
        1&\sin\phi\tan\theta&\cos\phi\tan\theta \\
        \dfrac{-uv}{u^2+w^2}&1&\dfrac{-vw}{u^2+w^2} \\[6pt]
        \dfrac{w}{\sqrt{u^2+w^2}}&0&\dfrac{-u}{\sqrt{u^2+w^2}} \\ 
        \end{bmatrix}
        \begin{bmatrix}
        p\\ 
        q\\ 
        r
        \end{bmatrix}
    \\
    &
    =
        \begin{bmatrix}
        0\\ 
        b_\alpha(\bm{x})\\ 
        b_\beta(\bm{x})
        \end{bmatrix}
        +
        \begin{bmatrix}
        \bm{a}_\phi(\bm{x})\\ 
        \bm{a}_\alpha(\bm{x})\\ 
        \bm{a}_\beta(\bm{x})
        \end{bmatrix}
		\bm{x}_2,
    \end{aligned}
    \label{eq:aero}
\end{equation}

\noindent where

\begin{equation}
\tilde{A}_x=A_x-g\sin\theta,\tilde{A}_y=A_y+g\sin\phi\cos\theta,\tilde{A}_z=A_z+g\cos\phi\cos\theta
\end{equation}

\noindent are the linear accelerometer measurements plus gravity components. The NDI control law for the aerodynamic angle loop is
\begin{equation}
	\bm{x}_{2,\mathrm{ref}}=
	\begin{bmatrix}
	p_{\mathrm{ref}}\\ 
	q_{\mathrm{ref}}\\ 
	r_{\mathrm{ref}}
	\end{bmatrix}
	=
	\begin{bmatrix}
	\bm{a}_\phi(\bm{x})\\ 
	\bm{a}_\alpha(\bm{x})\\ 
	\bm{a}_\beta(\bm{x})
	\end{bmatrix}^{-1}
	\left(
	\bm{\nu}_1
	-
	\begin{bmatrix}
	0\\ 
	b_\alpha(\bm{x})\\ 
	b_\beta(\bm{x})
	\end{bmatrix}
	\right).
\end{equation}

The dynamics of the inner loop in the incremental form are

\begin{equation}
	\dot{\bm{x}}_2 = 
    \dfrac{\mathrm{d}}{\mathrm{d}t}
    \begin{bmatrix}
    p\\ 
    q\\
    r
    \end{bmatrix}
    =
    \bm{J}^{-1}\left(
    \begin{bmatrix}
    M_x\\ 
    M_y\\ 
    M_z
    \end{bmatrix}
    -
    \begin{bmatrix}
    p\\ 
    q\\ 
    r
    \end{bmatrix}
    \times \bm{J}
    \begin{bmatrix}
    p\\ 
    q\\ 
    r
    \end{bmatrix}
    \right)
	\approx
	\begin{bmatrix}
	\dot{p}_0\\ 
	\dot{q}_0\\ 
	\dot{r}_0
	\end{bmatrix}
	+
	\bm{J}^{-1}\bm{J}_d^M(\bm{x}_0,\bm{\delta}_0)\Delta\bm{\delta},
\end{equation}

\noindent The INDI control law for the angular rate loop is

\begin{equation}
	\bm{x}_{3,\mathrm{ref}}=
	\bm{\delta}_{\mathrm{com}}
	=
	\left(\bm{J}_d^{M}\right)^{\dagger}\bm{J}
	\left(
	\bm{\nu}_2
	-
	\begin{bmatrix}
	\dot{p}_0\\ 
	\dot{q}_0\\ 
	\dot{r}_0
	\end{bmatrix}
	\right).
\end{equation}

\noindent It is assumed that the ICE aircraft is equipped with angular acceleration sensors to obtain measurements of angular rate derivatives directly. The control effectiveness matrix of aerodynamic moments $\bm{J}_d^{M}$ is updated at each time step by the onboard aerodynamic database developed in Ref.~\cite{RN26}. Note that $\bm{J}_d^{M}$ is non-square for the ICE aircraft due to over-actuation. The INCA scheme developed in Ref.~\cite{RN25} is applied to obtain a valid generalized inversion.

The throttle of the aircraft is controlled by a separated auto-throttle loop. The auto-throttle loop applies INDI to the following incremental dynamics of speed

\begin{equation}
\begin{aligned}
\dot{V}_g &= \frac{1}{m}\left(\cos\alpha\cos\beta\cdot F_x + \sin\beta\cdot F_y + \sin\alpha\cos\beta\cdot F_z\right) - g\sin\gamma \\
&\approx \dot{V}_{g,0} + \frac{1}{m}\left(\cos\alpha\cos\beta\cdot \dfrac{\partial F_x}{\partial T} + \sin\beta\cdot \dfrac{\partial F_y}{\partial T} + \sin\alpha\cos\beta\cdot \dfrac{\partial F_z}{\partial T}\right)\Delta T \\
&= \dot{V}_{g,0} + a_T(\bm{x})\Delta T,
\end{aligned}
\end{equation}

\noindent where the gradients of the aerodynamic forces depend on the MATV part only. The new thrust setting is selected as

\begin{equation}
T_{\mathrm{com}} = T_0 + \dfrac{k_T(V_{g,\mathrm{ref}}-V_g)-\dot{V}_{g,0}}{a_T(\bm{x})},
\end{equation}

\noindent where $k_T$ is the proportional gain. This dynamic inversion is not trivial for the ICE aircraft, since 1) the thrust force can be distributed in all three directions due to MATV, and 2) the ICE aircraft can operate at high AoA and sideslip angles.

\begin{figure}[hbt!]
	 \centering
	 \includegraphics[width=\linewidth]{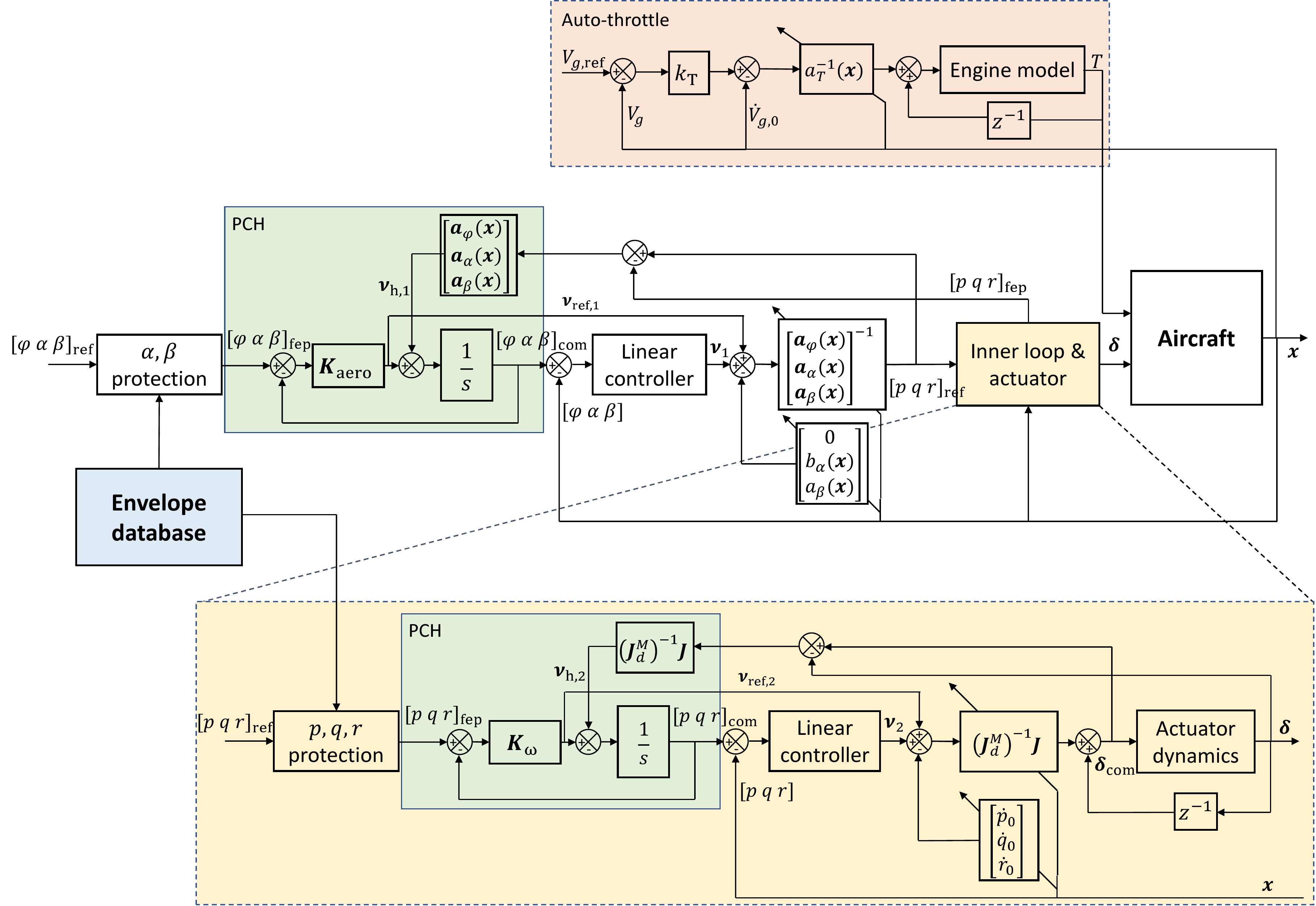}
	 \caption{Block diagram of FCS with FEP on the ICE aircraft.} 
	 \label{fig:block}
\end{figure}

To sum up, the complete block diagram of the envelope-protected nonlinear flight control system with aerodynamic angle commands is presented in Fig.~\ref{fig:block}.

\section{Results}
\label{sec:8}

\subsection{Estimated Envelope at Different Flight Conditions}

\begin{figure}[hbt!]
	 \centering
	 \includegraphics[width=\linewidth]{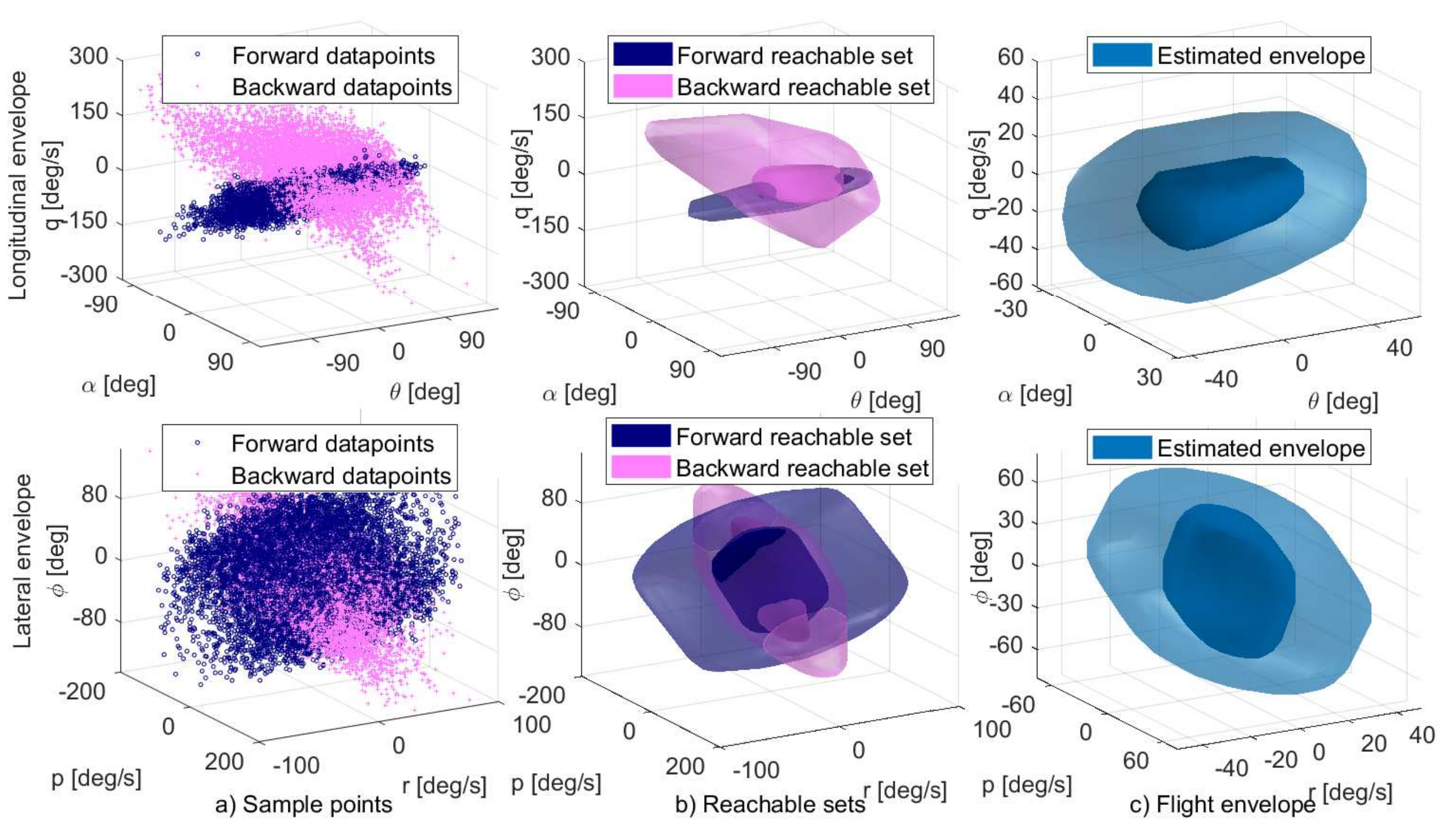}
	 \caption{Demonstration of envelope estimation with MC simulation under the baseline flight condition. Inner volume: the $\alpha$-cut at $1\sigma$, outer volume: the $\alpha$-cut at $2\sigma$.} 
	 \label{fig:env}
\end{figure}

First, the proposed probabilistic envelope estimation method as in Fig.~\ref{fig:mc_flow} is implemented under the baseline flight condition. The baseline flight operates at nominal mass configuration with full control effectors and a recovery time (time horizon) of 1.5s. The sample size of the MC simulation is 10,000. It is observed that the size and the shape of the estimated envelope stabilizes after this sample size under selected resolutions of the envelope database. The objective weighting factor in trimming $k_{\mathrm{trim}}$ is 1. Two 3-D slices of the estimated envelope are plotted in Fig.~\ref{fig:env} to show the longitudinal and lateral envelopes respectively at 20,000 ft and Mach 0.85. This routine starts with the raw MC simulation samples in Fig.~\ref{fig:env} a), followed by the kernel density estimators of reachable sets in Fig.~\ref{fig:env} b). It can be seen that the kernel density estimators can well capture the distribution of the samples. The probabilistic intersections of the reachable sets in Fig.~\ref{fig:env} c) form the estimated envelope. Two $\alpha$-cuts of the fuzzy sets at $k_0=1,2$ are plotted to demonstrate the probabilistic nature of the estimation.

\begin{figure}[hbt!]
     \centering
	 \subfloat[a) Recovery time.]{\includegraphics[width=\linewidth]{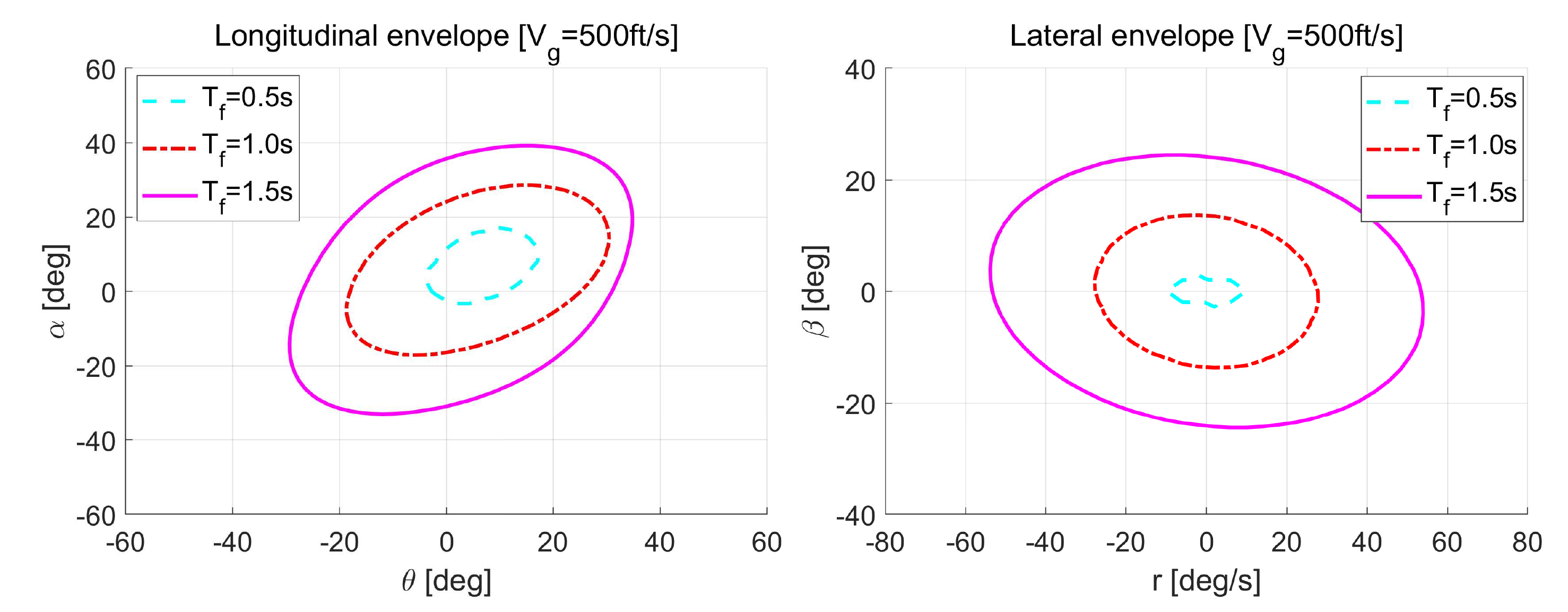}}\\
	 \subfloat[b) Thrust vectoring.]{\includegraphics[width=\linewidth]{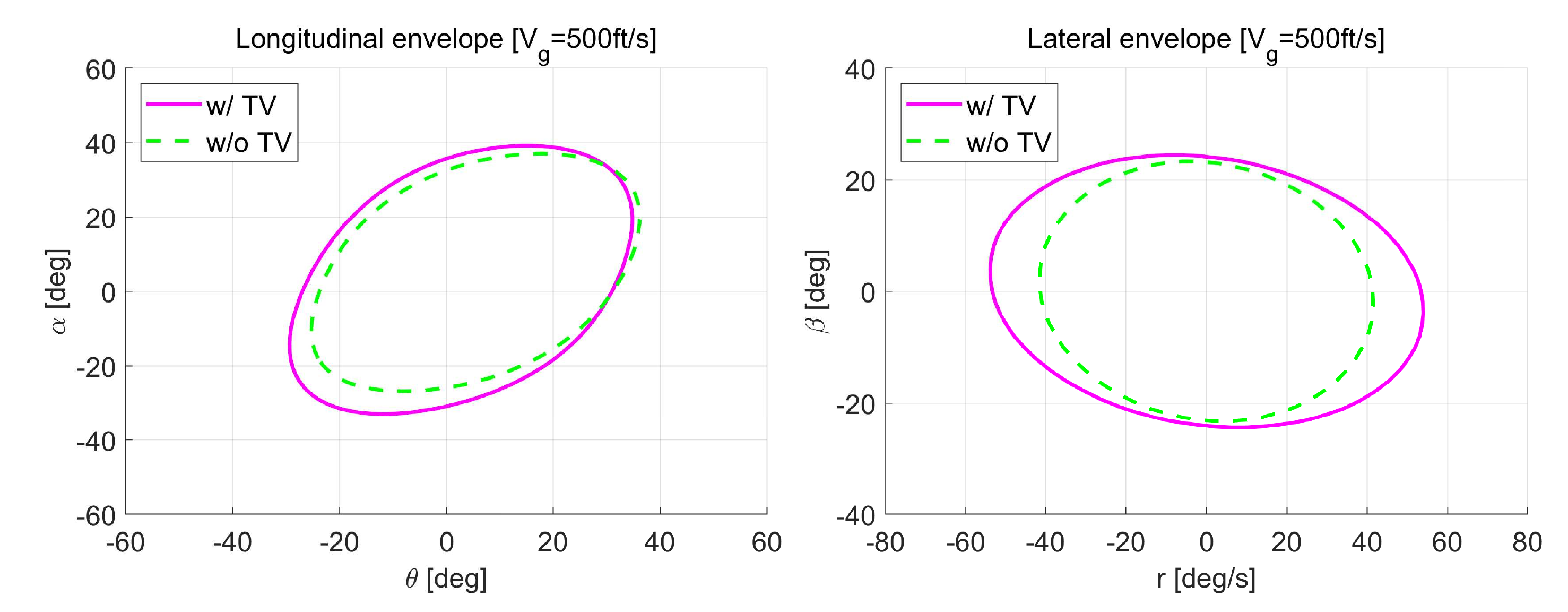}}\\
	 \subfloat[c) Mass configuration.]{\includegraphics[width=\linewidth]{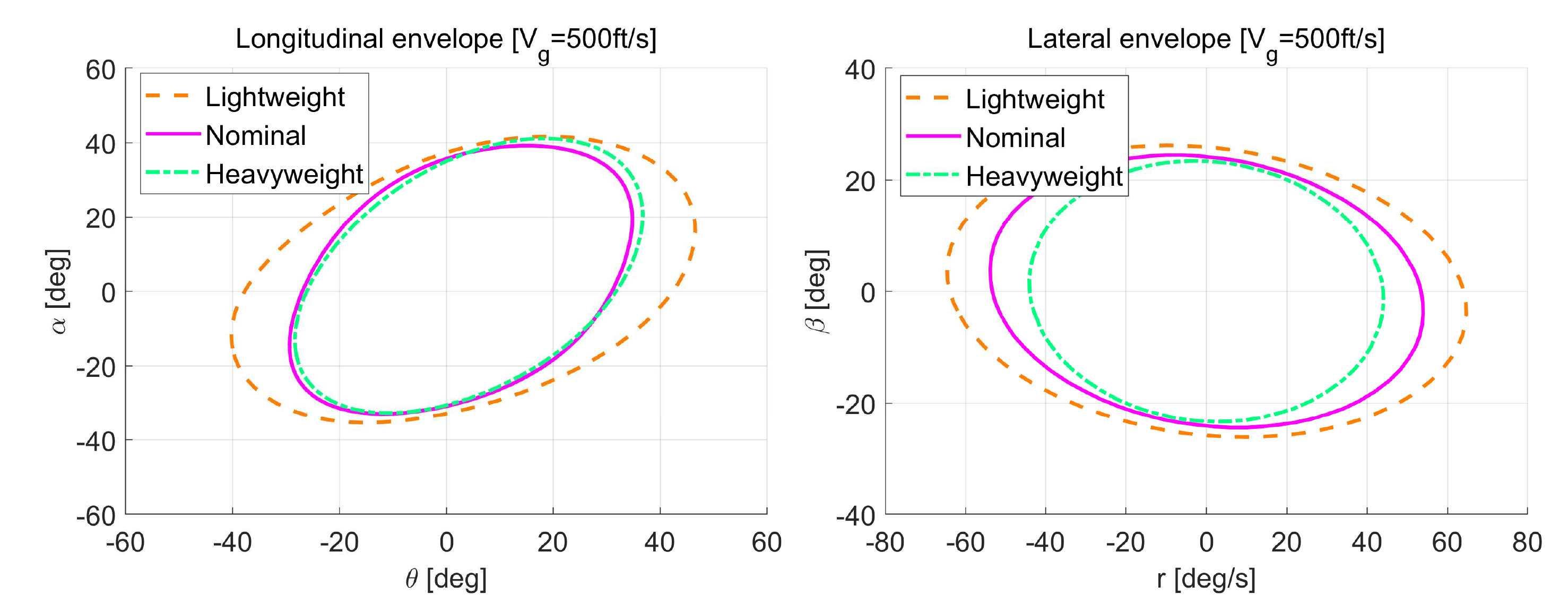}}
	 \caption{Comparison of envelope estimation results under different flight conditions.} 
	 \label{fig:comp}
\end{figure}

Since no other estimation methodology is tractable for this seven-dimensional non-linear problem, direct comparison of estimation results is not possible. Instead, to show the estimation gives reasonable results, it is compared under different flight conditions, namely recovery time, equipped control effectors, and mass configurations. Fig.~\ref{fig:comp} shows 2-D slices of different estimated envelopes at 20000 ft and 500 ft/s, longitudinally with respect to $\alpha$ and $\theta$ and laterally with respect to $\beta$ and $r$. The recovery time (Fig.~\ref{fig:comp} a)) is the main factor that affects the size of the envelope since it limits the length of all trajectories. However, this universal parameter has limited effects on the shape of the envelope. In terms of equipped control effectors, we demonstrated the effect of thrust vectoring as it can exhibit directionality by mainly augmenting the controllability in the yaw direction. As observed in Fig.~\ref{fig:comp} b), this directional effector has a big effect on the shape of the envelope, particularly enlarging the envelope in the yaw direction. In Fig.~\ref{fig:comp} c), a lighter mass configuration enhances the agility of the aircraft and thus leads to a larger envelope. The shape of the longitudinal envelope is also altered due to the shift of the center of gravity under different mass configurations.

The above observations indirectly offer some insights on the validity of the estimation method. The following section continues to demonstrate its effectiveness together with the FEP-equipped FCS by case study.

\subsection{Case Study of LOC Avoidance Capability}

The performance of the envelope-protected controller was tested by two sample maneuvers. These maneuvers are selected such that LOC is observed with a baseline controller without FEP. The first command, known as Maneuver A, is a triangular $\alpha$ signal with a base of 4 s and a height of 90 deg. The selection of triangular signals over pulses or doublets is to demonstrate the difference between the two FEP strategies: state-constraint-based FEP only activates above a certain threshold, whereas probabilistic FEP activates throughout the flight. The second command, known as Maneuver B, is a combined $\alpha$-$\beta$ command of two consecutive triangular signals with opposite signs. The triangular signals have a base of 6 s and a height of 50 deg. The $\beta$ signal lags 2 s behind the $\alpha$ signal. Both commands are prefiltered by a low-pass filter.

The ranges and resolutions of the states stored in the envelope database are shown in Table~\ref{tbl:range}. The applied linear controller gains for both loops and the auto-throttle are listed in Table~\ref{tbl:gain}, as well as the compensation gains of the probabilistic FEP $\bm{K}_{\mathrm{fep}}$. The reference model gains in PCH $\bm{K}_{\mathrm{ref},i}$ were selected the same as the proportional gains of the same loop $\bm{K}_{\mathrm{p},i}$. The threshold to activate the probabilistic envelope protection $M_0$ is 0. For the state-constraint-based FEP, the probabilistic envelope is binarized with a threshold setting of $k_0=3$. All simulations of the controller start with a trimmed flight at 20,000 ft and Mach 0.85 and full control effectors.

\begin{table}
\parbox{.40\linewidth}{
\centering
\caption{Grid of the envelope database}
\label{tbl:range}
\renewcommand{\arraystretch}{1.1}
\begin{tabular}{@{}cccc@{}}
\midrule\midrule
                   & \textbf{Min} & \textbf{Max} & \textbf{Step size} \\ \midrule
$p$ {[}deg/s{]}    & -150         & 150          & 30           \\
$q$ {[}deg/s{]}    & -150         & 150          & 30           \\
$r$ {[}deg/s{]}    & -60          & 60           & 30           \\
$\alpha$ {[}deg{]} & -60          & 60           & 5            \\
$\beta$ {[}deg{]}  & -45          & 45           & 5            \\
$V_g$ {[}ft/s{]}   & 400          & 1300         & 180          \\
$h$ {[}ft{]}       & 10,000       & 30,000       & 5000         \\ \midrule\midrule
\end{tabular}
}
\hfill
\parbox{.60\linewidth}{
\centering
\caption{Summary of controller gains.}
\label{tbl:gain}
\renewcommand{\arraystretch}{1}
\begin{tabular}{@{}cccccccc@{}}
\midrule\midrule
                        & \multicolumn{3}{c}{\textbf{Inner loop}} & \multicolumn{3}{c}{\textbf{Outer loop}} & \textbf{Auto-throttle} \\ \cmidrule(l){2-8} 
                        & $p$         & $q$         & $r$         & $\phi$     & $\alpha$     & $\beta$     & $T$                    \\ \midrule
$\bm{K}_{\mathrm{p}}$   & 6.50        & 6.50        & 5.80        & 2.00       & 2.00         & 1.60        & 1.00                   \\
$\bm{K}_{\mathrm{i}}$   & 0.00        & 0.00        & 0.00        & 0.50       & 0.50         & 0.30        & 0.00                   \\
$\bm{K}_{\mathrm{d}}$   & 0.00        & 0.00        & 0.50        & 0.90       & 0.90         & 0.00        & 0.00                   \\
$\bm{K}_{\mathrm{fep}}$ & 0.05        & 0.05        & 0.05        & N/A        & 0.80         & 0.80        & N/A                    \\ \midrule\midrule
\end{tabular}
}
\end{table}

\begin{figure}[hbt!]
   \centering
	 \subfloat[a) Maneuver A.]{\includegraphics[width=3.25in]{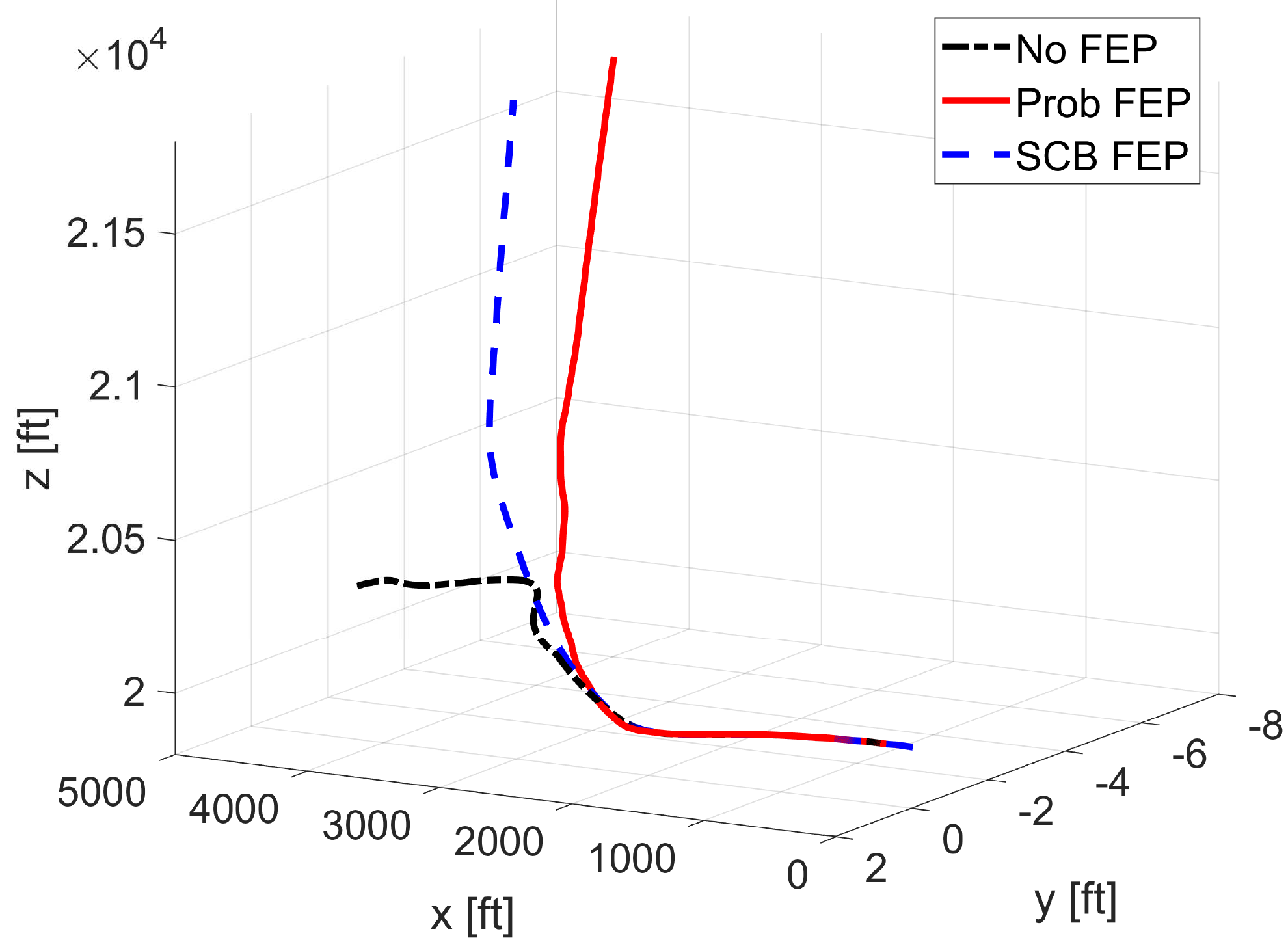}}
	 \subfloat[b) Maneuver B.]{\includegraphics[width=3.25in]{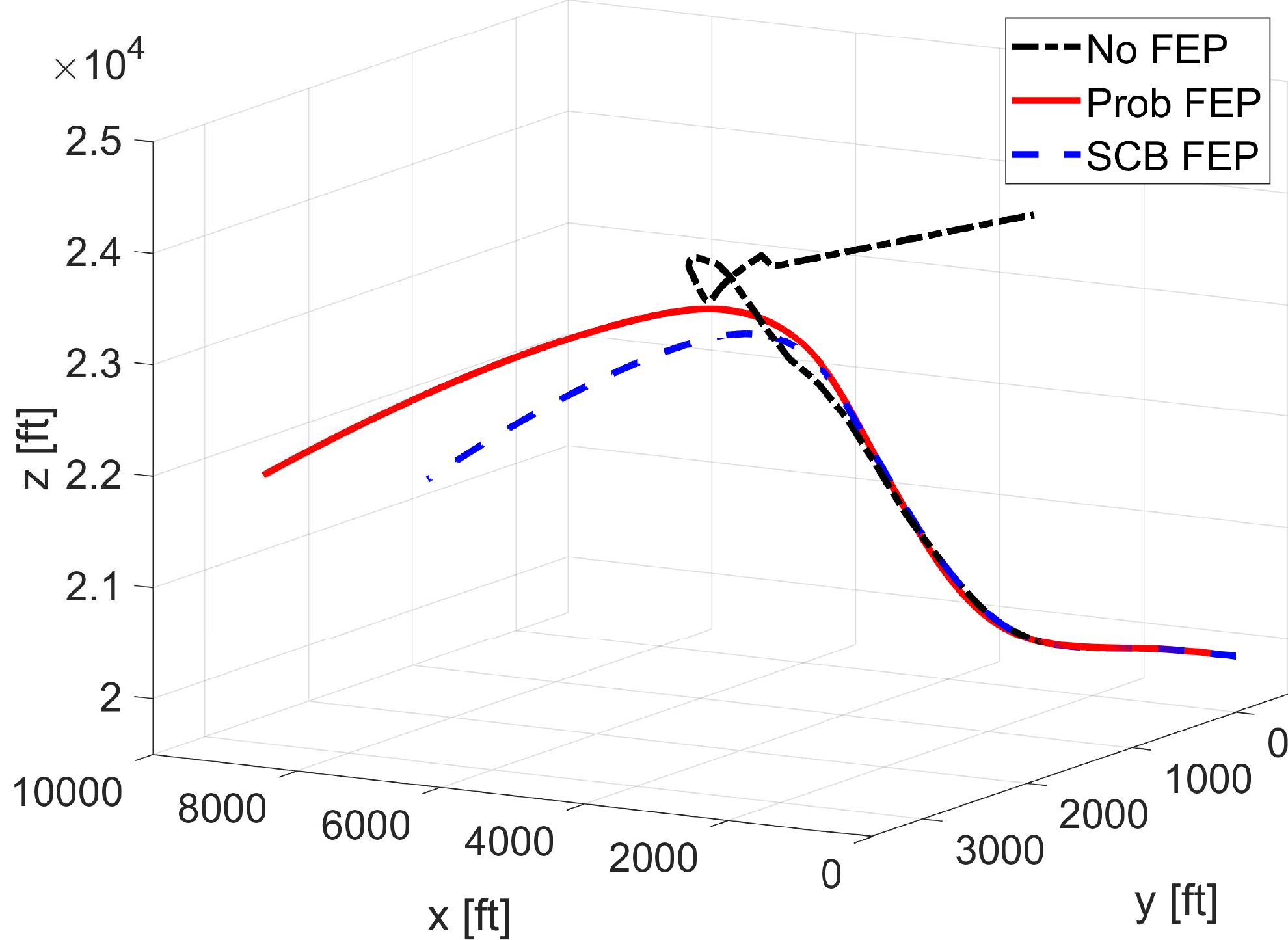}}
	 \caption{3-D trajectories of Maneuver A and B with different FEP settings. Initial position: [0, 0, 20000] ft} 
	 \label{fig:3dtraj}
\end{figure}

\begin{figure}[hbt!]
	 \centering
	 \subfloat[a) Maneuver A.]{\includegraphics[width=3.25in]{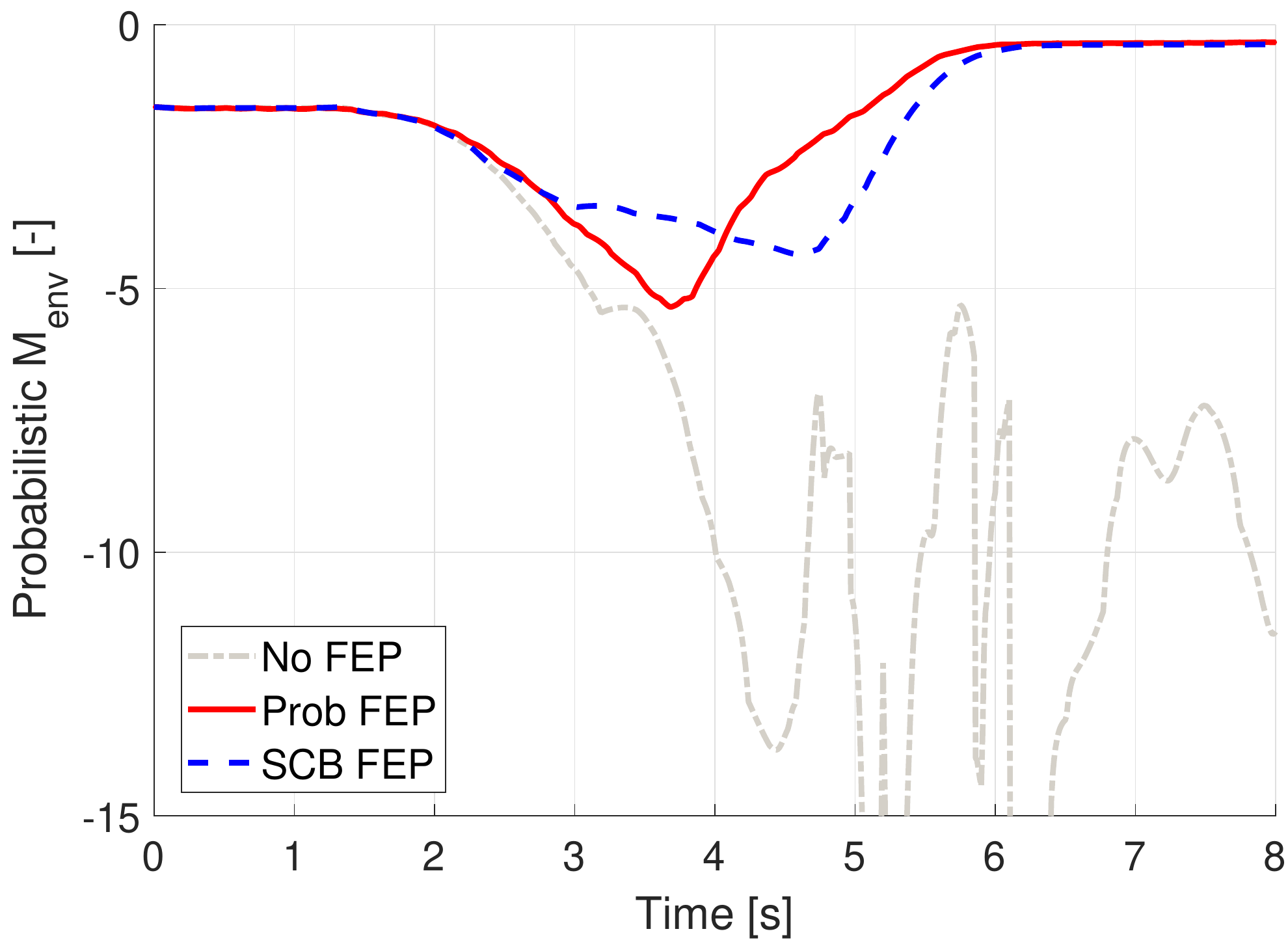}}
	 \subfloat[b) Maneuver B.]{\includegraphics[width=3.25in]{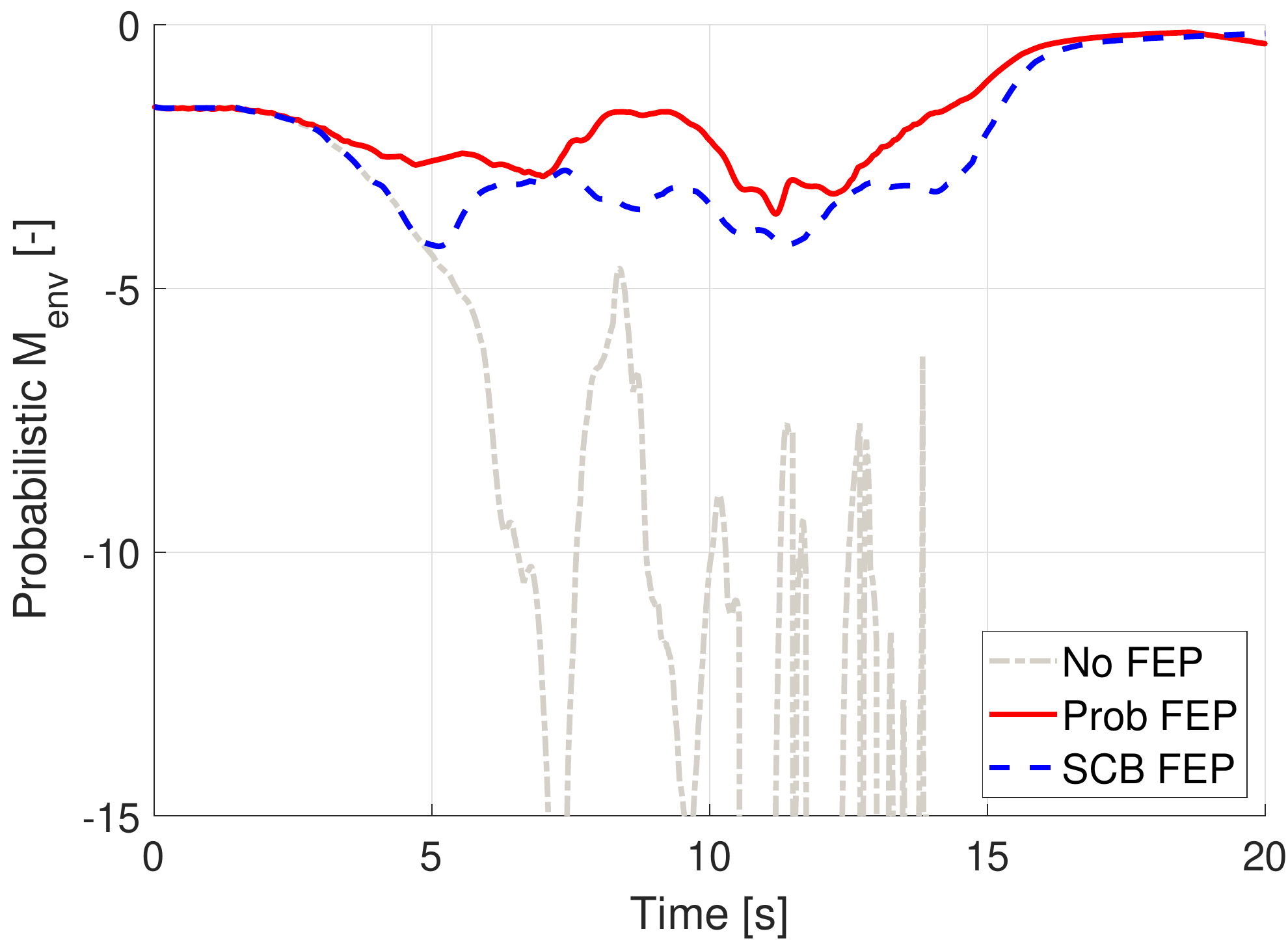}}
	 \caption{Effectiveness of FEP in terms of the probabilistic envelope metric.} 
	 \label{fig:metric}
\end{figure}

These two commands are simulated with three different controllers: no FEP, with probabilistic FEP (PROB-FEP), and with state-constraint-based FEP (SCB-FEP). 3-D trajectories of the simulated flights are plotted in Fig.~\ref{fig:3dtraj}. It can be seen that for both maneuvers, LOC is observed when no FEP is applied, which leads to severely uncontrolled flights. A direct comparison of the FEP capability is shown in Fig.~\ref{fig:metric} by comparing the probabilistic envelope metric. A higher value means the aircraft is safer in the envelope. So the general conclusion is that both FEP strategies can constrain the aircraft within the safe part of the state space and avoid LOC.

\begin{figure}[hbt!]
   \centering
	 \subfloat[a) State-constraint-based FEP.]{\includegraphics[width=3.25in]{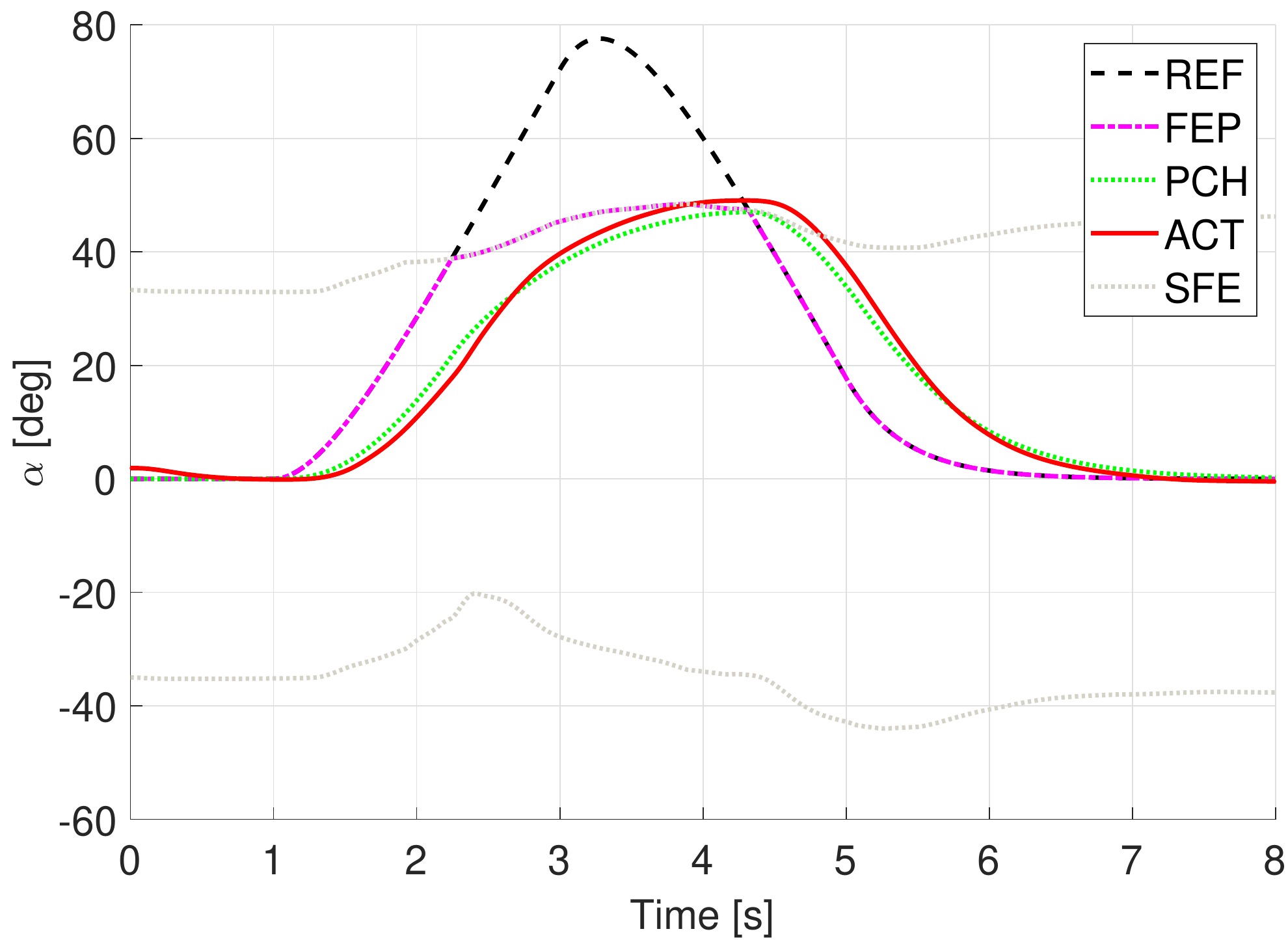}}
	 \subfloat[b) Probabilistic FEP.]{\includegraphics[width=3.25in]{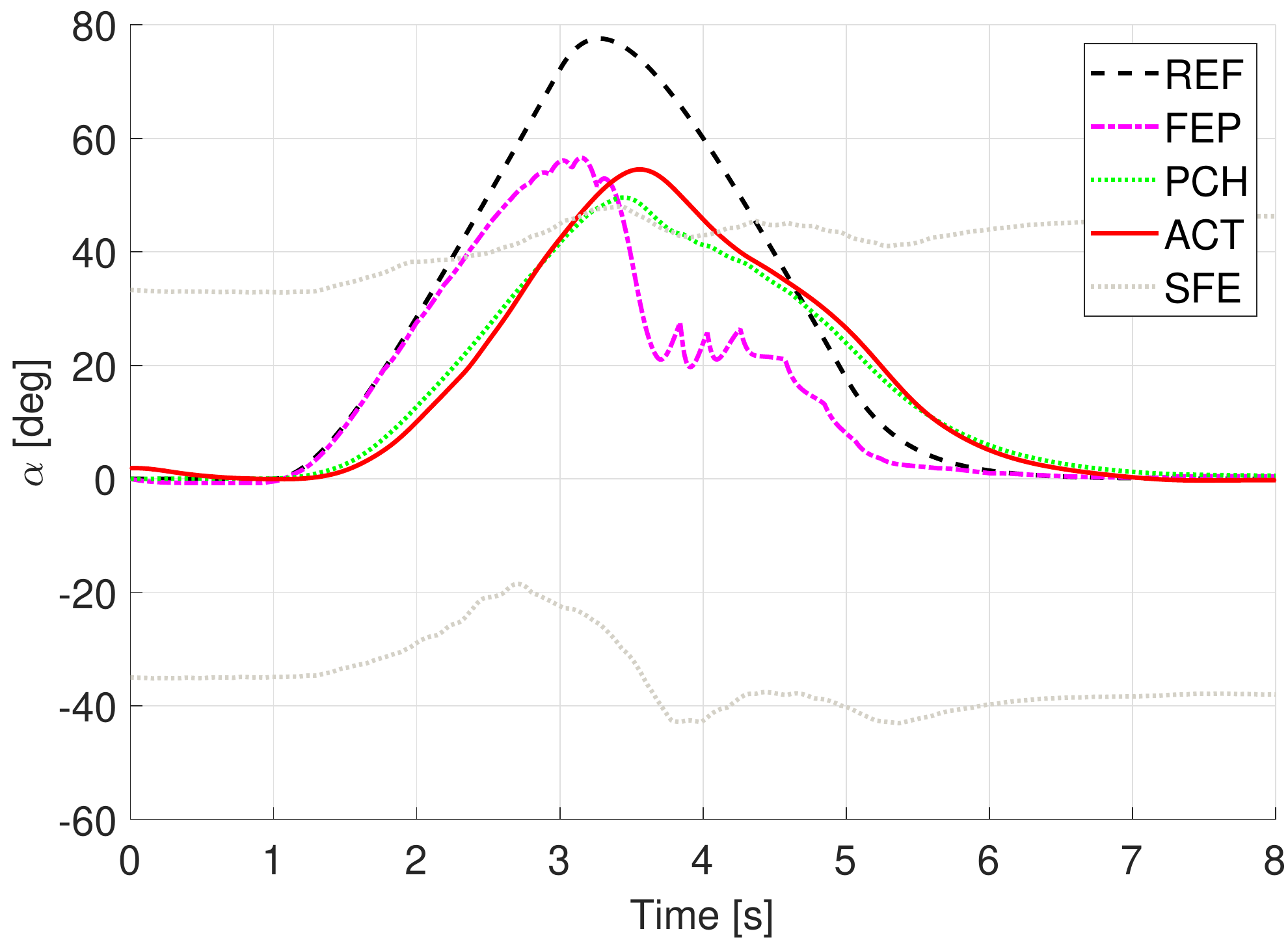}}
	 \caption{Comparison of FEP performances for Maneuver A. REF: reference, FEP: command after FEP, PCH: command after PCH, ACT: actual response, SFE: estimated envelope boundary.} 
	 \label{fig:maneua}
\end{figure}

\begin{figure}[hbt!]
     \centering
	 \subfloat[a) State-constraint-based FEP.]{\includegraphics[width=3.25in]{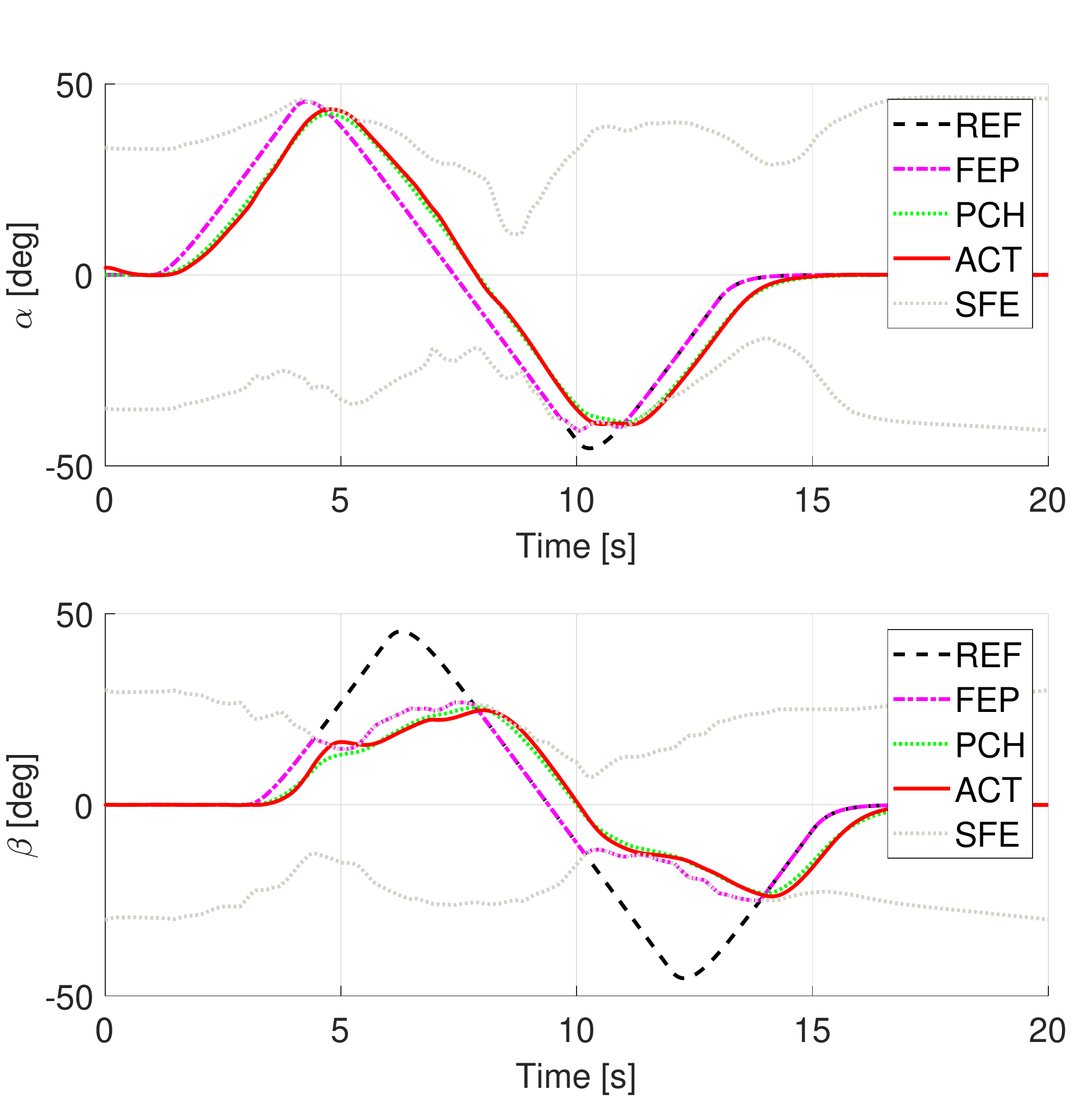}}
	 \subfloat[b) Probabilistic FEP.]{\includegraphics[width=3.25in]{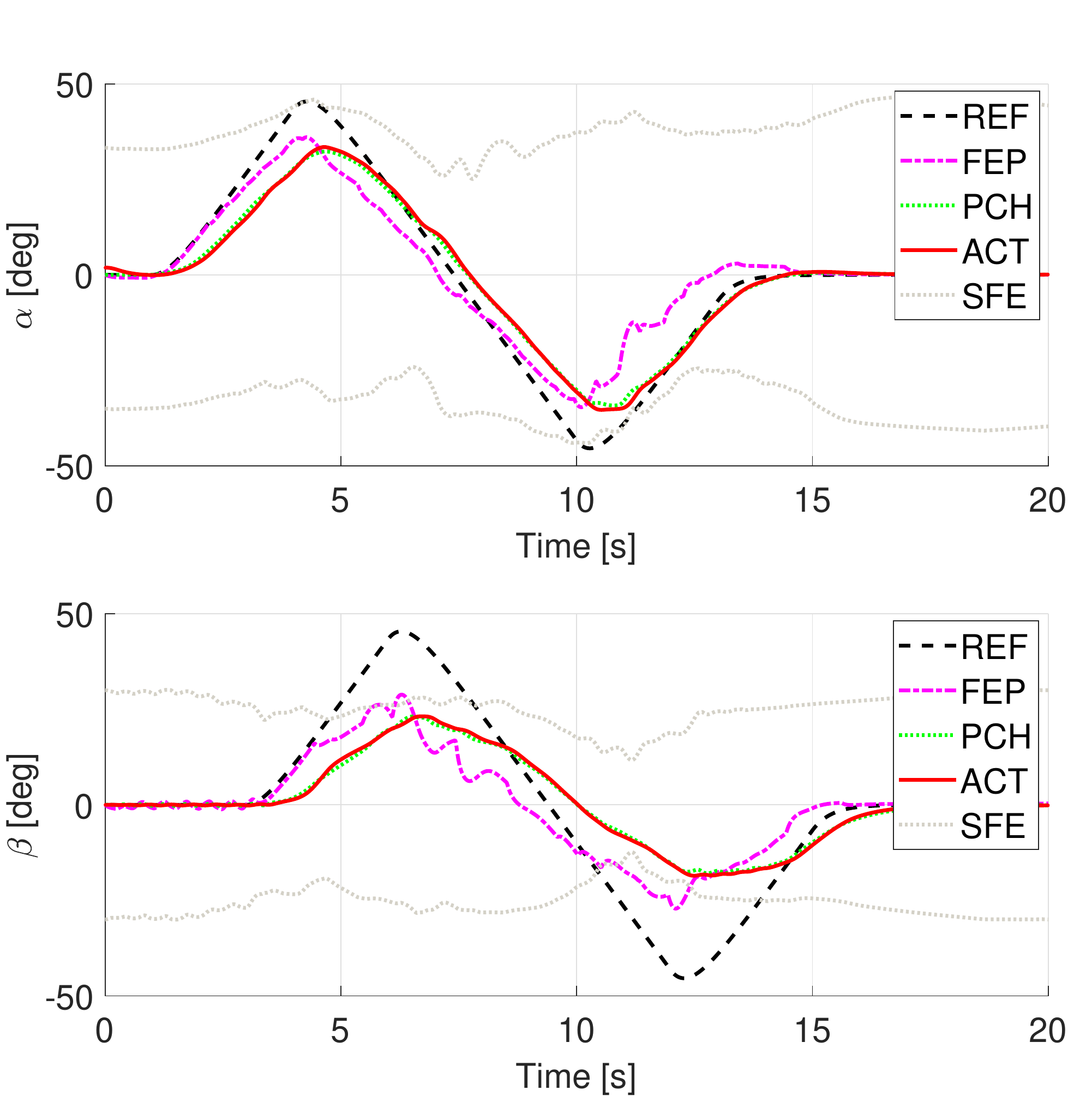}}
	 \caption{Comparison of FEP performances for Maneuver B. REF: reference, FEP: command after FEP, PCH: command after PCH, ACT: actual response, SFE: estimated envelope boundary.} 
	 \label{fig:maneub}
\end{figure}

The detailed controller performances are shown in Fig.~\ref{fig:maneua} and Fig.~\ref{fig:maneub} for Maneuver A and B respectively. Despite their overall success, they behave differently in face of dangerous commands. SCB-FEP performs exactly the same as the unprotected controller in the beginning. Then after detection of (absolute) dangerous commands, the FEP system gives protected commands that strictly follows the estimated state constraints immediately. In contrast, PROB-FEP actively modifies the commands throughout the flight based on the membership value of the current state but with no fixed boundaries. As a result, SCB-FEP gives trapezoid responses by cutting off the top of the triangular commands, whereas PROB-FEP better preserves the command shape with responses of smaller magnitudes. Another difference observed in Maneuver B (Fig.~\ref{fig:maneub}) is that the protection of different states is isolated in SCB-FEP but integrated in PROB-FEP. In SCB-FEP, the AoA command is not protected as it does not violate the binary state constraint, whereas in PROB-FEP, it is also protected as it helps to enlarge the sideslip angle envelope.

\section{Conclusion}
\label{sec:9}

This paper presents a novel framework to apply high-dimensional envelope estimation and protection to aircraft with complex nonlinear models. The framework extends the definition of the flight envelope to fuzzy sets and uses Monte Carlo simulation with extreme control effectiveness sampling to circumvent solving the optimization problem directly. In this way, the probabilistic envelope can be estimated conservatively and efficiently under flexible flight conditions for no less than seven dimensions. Furthermore, the estimated envelope is stored onboard to provide online dynamic flight envelope protection to a multi-loop nonlinear dynamic inversion controller with both conventional command limiting and novel probabilistic modification strategies. The command limiting strategy guarantees a fixed command region, whereas the probabilistic strategy offers predictive, multivariate, and shape-preserved protection. Simulation results on a high-performance fighter aircraft show both good validity of the estimation and effective protection capability to avoid loss of control with both strategies.

\bibliography{Bibliography}

\end{document}